\documentclass[10pt,journal]{IEEEtran} %

\usepackage[]{cite} %
\usepackage[pdftex]{graphicx} %
\graphicspath{{fig/}} 
\usepackage{amsmath}
\usepackage{array}
\usepackage[caption=false,font=footnotesize]{subfig}

\usepackage{bbm} %
\usepackage{amssymb} %
\usepackage{booktabs} %
\usepackage{tabularx} %
\usepackage{multirow} %
\usepackage{tikz} %
\usetikzlibrary{positioning}
\usetikzlibrary{shapes,arrows,shadows}
\usetikzlibrary{decorations.pathreplacing}
\usepackage{adjustbox}
\usepackage{amsthm} %
\usepackage{enumitem}   %
\usepackage{xcolor}
\usepackage{soul}

\usepackage[flushleft]{threeparttable} %

\usepackage[ruled,linesnumbered]{algorithm2e} %
\usepackage[hang,flushmargin]{footmisc}
\usepackage{lipsum}
\makeatletter
\newcommand{\algorithmfootnote}[2][\footnotesize]{%
  \let\old@algocf@finish\@algocf@finish%
  \def\@algocf@finish{\old@algocf@finish%
    \leavevmode\rlap{\begin{minipage}{\linewidth}
    #1#2
    \end{minipage}}%
  }%
}
\makeatother

\newtheorem{definition}{Definition}
\newtheorem{proposition}{Proposition}

\begin{document}

\title{Mitigating Traffic Remapping Attacks in Autonomous Multi-hop Wireless Networks}

\author{Jerzy~Konorski
        and~Szymon~Szott%
\thanks{J. Konorski is with the Gdansk University of Technology, Gdansk, Poland, e-mail: jekon@eti.pg.edu.pl.}%
\thanks{S. Szott is with the AGH University of Science and Technology, 
Krakow, Poland, e-mail: szott@agh.edu.pl.}%
}

\maketitle

\begin{abstract}
Multi-hop wireless networks with autonomous nodes are susceptible to selfish traffic remapping attacks (TRAs). Nodes launching TRAs leverage the underlying channel access function to receive unduly high quality of service (QoS) for packet flows traversing source-to-destination routes.
TRAs are easy to execute, impossible to prevent, difficult to detect, and harmful to the QoS of honest nodes. 
Recognizing the need for providing QoS security, we use
a novel network-oriented QoS metric to propose a self-enforcing game-theoretic mitigation approach.
By switching between TRA and honest behavior, selfish nodes engage in a noncooperative multistage game in pursuit of high QoS. We analyze feasible node strategies and design a distributed signaling mechanism called DISTRESS, under which, given certain conditions, the game produces a desirable outcome: after an upper-bounded play time, honesty tends to become a selfish node's best-reply behavior, while yielding acceptable QoS to most or all nodes.
We verify these findings by Monte Carlo and ns-3 simulations of static and mobile nodes.
\end{abstract}

\begin{IEEEkeywords}
Wireless networks, autonomous nodes, quality of service security, selfish attacks, game theory, IEEE 802.11 
\end{IEEEkeywords}

\section{Introduction}\label{sec:introduction}

\IEEEPARstart{M}{ultihop} wireless transmission is the underlying principle of various types of wireless networks: sensor, mesh, ad hoc, cellular (using cooperative user-to-base-station relays), vehicular, opportunistic (delay tolerant) and Internet of Things (IoT) segments.
In this paper we focus on wireless networks with autonomous nodes, referred to as \emph{multi-hop autonomous wireless networks} (MAWiNs), which are 
an active research field exploring the self-organizing network concept \cite{Atayero2014} and its numerous embodiments, such as autonomous mobile mesh networks \cite{Haseeb2020, Pirmagomedov2020}, 
autonomous vehicular networks \cite{Peng2020}, 
autonomous sensor networks \cite{Rodriguez2020}, and
flying ad hoc networks \cite{srivastava2021future}.
In large-scale IoT systems, multihop cooperative relaying has been observed to improve the throughput, reliability, and energy efficiency of the data exchange between end-user devices (e.g., smart meters) and gateway nodes (e.g., data aggregation points) \cite{bader2016localized,omar2017experimental,ramezan2018survey,liu2020index,chen2019performance}. 

Besides classical threats to wireless transmission, MAWiNs face unique security threats due to node autonomy. Sustained operation of such networks relies on the nodes' benevolent compliance with cooperative protocols, such as fair channel access, transit packet forwarding on behalf of out-of-range source and destination nodes, and route discovery. Nodal cooperation entails certain costs in terms of energy expenditure and quality of service (QoS) received by source (locally generated) traffic: fair channel access requires deferment of one's own packet transmission, which causes delays; transit packet forwarding consumes energy and bandwidth, which diminishes source traffic throughput; and route discovery introduces communication overhead and processing burden. 
Autonomous nodes thus tend to exhibit rational (selfish) behavior, seeking a favorable tradeoff between received QoS and incurred costs. This makes MAWiNs susceptible to selfish attacks that abuse the employed network protocols to the attacker's benefit.

Selfish attacks
can be categorized as ``supply-side'' (attempts to reduce costs of performing network services for other nodes) and ``demand-side'' (attempts to acquire undue network resources through aggressive competition). The former are common at the Internet layer, e.g., refusal to forward transit packet saves energy and bandwidth for source traffic, and falsifying route advertisements may deactivate incident routes and thus relieve a node from transit traffic. In response, defense mechanisms need to be deployed, such as secure routing protocols \cite{Kannhavong2007}, intrusion detection/prevention systems \cite{Khan2020}, credit-based schemes \cite{Samian2015}, and trust management frameworks \cite{Movahedi2016}. 

Known ``demand-side'' selfish attacks range from the link layer to the transport layer and usually consist in manipulation of sensitive protocol parameters, e.g., the contention window of IEEE 802.11 \cite{80211-2016} or congestion control settings of TCP \cite{ rfc5681}. They have been countered by a number of detection- or prevention-type mechanisms \cite{Li2015} and often investigated using game theory \cite{Akella2002}, e.g., for incentivizing selfish IoT devices to participate in cooperative communication \cite{afghah2018reputation,zhang2019incentivizing}.
A less known variety of selfish attacks by the name \emph{QoS abuse} \cite{Szigeti2013} or \emph{traffic remapping attack} (TRA) \cite{Szott2017} emerges in environments supporting traffic class-based QoS differentiation to enforce user-network QoS contracts such as Service Level Agreements (SLAs) \cite{rfc3198}.
By falsely assigning traffic to classes, an attacker node abuses provisioned QoS policies and can receive a higher QoS level for its source traffic at the cost of honest nodes' source traffic. 
MAWiNs make a perfect scene for TRAs especially if they offer QoS differentiation as part of the network's mission rather than a metered service, which makes launching a TRA costless, whereas their inherent lack of node accountability reduces the risk of detection and punishment. Thus on top of classical problems with network resource deficiency and/or mismanagement, threats to contractual QoS can arise from QoS abuse and should be addressed by a new class of defense mechanisms that can be collectively termed \emph{QoS security}. Their task is to protect information security, in the sense of enforcing QoS guaranteed to honest nodes, in the presence of QoS abuse.

\begin{figure}[t]
\centering
\includegraphics[width=\columnwidth]
{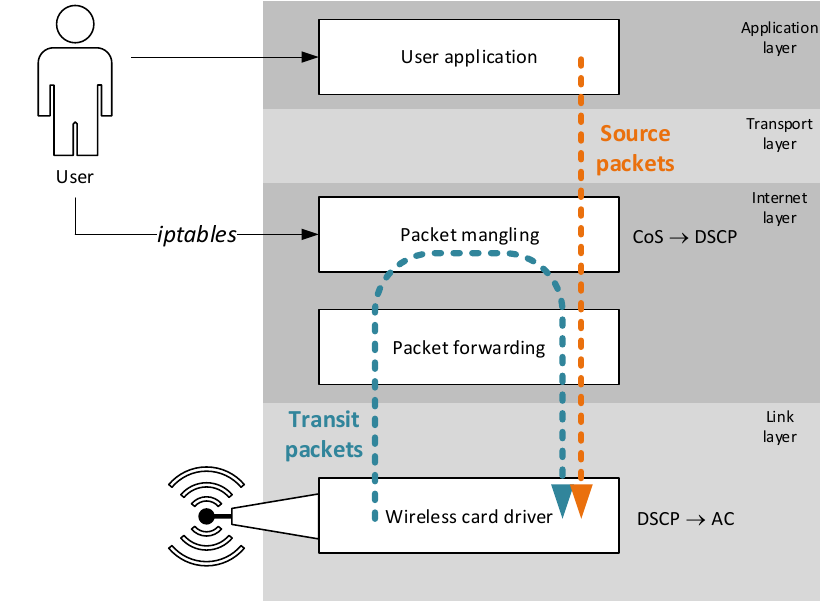}%
\caption{Packet flow and TRA execution at an attacker node in the TCP/IP model.}
\label{fig:tra-stack}
\end{figure}

In wireless networks using IEEE 802.11, TRAs can exploit the enhanced distributed channel access (EDCA) function. 
EDCA defines four access categories (ACs), each with its own parameters controlling the priority and duration of medium access \cite{80211-2016}. 
Packets are mapped to ACs based on the Differentiated Services Code Point (DSCP) in their IP header, which reflects the traffic's Class of Service (CoS) \cite{Stankiewicz2011}.
For simplicity assume that DSCP can be either \emph{expedited forwarding} (EF) or \emph{best effort} (BE). The CoS-to-DSCP mapping is implemented by Internet layer packet mangling software, such as Linux \texttt{iptables}. TRAs can be easily executed by setting DSCP = EF in source traffic whose CoS maps to BE %
and DSCP = BE in forwarded (transit) traffic whose CoS maps to EF,
cf. Fig.~\ref{fig:tra-stack}. 
In contrast, AC parameter modification requires tampering with the DSCP-to-AC mapping embedded in wireless card drivers.
Furthermore, TRAs are difficult to detect: determining if higher-layer traffic matches its DSCP designation requires deep packet inspection \cite{Konorski2014} and global knowledge of the DSCP assignment policy.

In single-hop settings, TRAs have been
shown to drastically reduce the throughput of honest nodes unless the latter employ a carefully designed MAC-layer discouragement scheme \cite{Konorski2014}. The multi-hop nature of MAWiNs poses a number of additional challenges for a defense scheme:
 
\begin{itemize}
    \item a selfish node can both promote its source traffic and demote transit traffic,
    \item a locally performed TRA has an end-to-end impact: once assigned a false QoS designation, a packet retains it further down the route,
    \item the impact of TRA is unclear \emph{ex ante}, due to the complex interplay of multiple layers: PHY (hidden nodes), MAC (channel contention), and transport (flow control); this interplay also blurs QoS perception and
    rules out straightforward detection-based countermeasures,
    \item the absence of single-broadcast hearability rules out simple punitive measures such as threats of jamming,
    \item heuristic end-to-end countermeasures against TRAs are moderately effective \cite{Szott2017}; in particular, countermeasures that work well in single-hop settings, such as ACK dropping, fail in multi-hop settings \cite{Szott2018}.
\end{itemize}

Being rationally motivated, easy to execute, impossible to prevent, difficult to detect, and harmful to honest nodes, TRAs call for incentive-based defense. Unfortunately, the lack of a central authority rules out common approaches based on reputation building or Stackelberg games \cite{bader2016localized,omar2017experimental,Movahedi2016}, hence a novel game-theoretic methodology is also needed.
In \cite{Konorski2017a}, we presented an early formulation of the multihop TRA problem and preliminary insights into its impact upon the nodes’ cost metric. A multistage game-type TRA mitigation scheme was proposed, whose convergence and alignment with nodes' rationality was only supported by numerical and Monte Carlo simulation arguments.
Building on a more rigorous network model
we offer herein a provably convergent, rational, and effective TRA mitigation scheme in the spirit of ``brinkmanship game theory'' \cite{Jahan2010}: credible threats force nodes to toggle between TRA and honest behavior, and so engage in a noncooperative game in pursuit of high QoS. Although our analytical results apply to networks with static nodes, simulations show the scheme is also effective with node mobility.
Our main contributions are:

\begin{enumerate}
\item Based on a MAWiN model with a static topology and traffic flows,
we formally define \emph{plausible opportunistic} TRAs, and discuss their motivation and impact.
\item We develop a heuristic end-to-end QoS metric that only uses information about the network topology and traffic flows, and verify it by simulation and comparison with alternative heuristics.
\item We design a distributed DISTRESS mechanism to signal the threat of service suspension due to ongoing TRAs. DISTRESS requires little data analysis and inter-node synchronization, and needs not distinguish between TRA and objectively harsh traffic conditions.
\item Using the developed QoS metric as a payoff function,
we analyze the game arising among ill-behaved nodes under DISTRESS and state conditions of its desirable outcome. We show that after an upper-bounded play time, honesty tends to become an ill-behaved node's best-reply behavior, keeping QoS acceptable to most or all nodes. These findings are verified by extensive Monte Carlo and time-true simulations of static and mobile nodes.
\end{enumerate}

The remainder of the paper is organized as follows. In Section~\ref{sec:soa} we outline related work on QoS abuse in wireless networks and highlight the unsolved problems which justify our research, including the need for a macroscopic MAWiN network model. In Section~\ref{sec:topology} we formulate a topology and traffic flow model, next used in Section~\ref{sec:attacks} to formalize the notion of TRA. In Section~\ref{sec:performance} we develop a MAWiN performance model and propose an end-to-end QoS metric; the latter is shown in Section~\ref{sec:motivation} to yield quantitative insight into the motivation and impact of TRAs. In Sections~\ref{sec:one-shot-game} and \ref{sec:multistage-game}, respectively, we describe the one-shot TRA game arising among ill-behaved nodes, and propose a model of multistage play to discourage TRAs. ``Good'' multistage strategies are analyzed in Section~\ref{sec:multistage-strategy} and validated by simulations in Section~\ref{sec:simulations}. Section~\ref{sec:conclusions} concludes the paper.

\section{Related Work}
\label{sec:soa}

Selfish attacks in MAWiNs have mostly been studied at the Internet layer. The main attack under consideration has been packet dropping, also referred to as \emph{forwarding}/\emph{relaying misbehavior}. This attack can be considered as launched either on all packets (full dropping) or only on selected packets (partial dropping). In the latter case the dropping can be either probabilistic or deterministic (e.g., may specifically target some packet types such as routing control packets, or some source-to-destination routes). 

The packet dropping attack has been widely analyzed. Due to node autonomy and lack of any administrative control only ``soft'' countermeasures are possible. Some proposals involve micropayment (credit) schemes, where a virtual currency is earned for relaying and next used to buy similar services \cite{Samian2015}. %
Others have focused on explicit identification of attackers. This can be done passively, e.g., through a watchdog mechanism where nodes promiscuously listen to the channel and observe offending behavior \cite{Ren2016}, %
or actively, e.g., using additional end-to-end acknowledgments to determine which routes contain packet dropping attackers \cite{Liu2007}. %
Attacker identification can be enhanced through complex audit- and reputation-based schemes where a node derives a reputation score of any other node from first-hand (watchdog-based) experience, and possibly from reputation scores calculated by third-party nodes; low-reputation nodes are identified as attackers \cite{Zhang2016}. %

The main response to Internet layer attacks has been of a reciprocation nature, i.e., restricted forwarding of attackers' source packets. This gives rise to numerous analyses of
the underlying forwarding game. Various strategies (e.g., tit-for-tat) have been considered to enforce honest packet forwarding; see \cite{Liu2010} for a systematic treatment.  %
A complementary response is to route traffic around attackers. However, this is in fact beneficial to them, as they can expend less energy and bandwidth on forwarding \cite{Ayday2010}.  %

At the link layer, most attacks have found the IEEE 802.11 channel access function \cite{80211-2016} an easy target. Numerous studies have shown that launching a \emph{backoff attack}, i.e., changing the transmission deferment parameters (such as idle carrier sensing or backoff times) yields the attacker a considerable increase in throughput and access delays at the cost of honest nodes \cite{Parras2018,fihri2020machine}. %
Link layer attacks have mostly been studied in a single-hop setting, which is not surprising given their local-scope nature. 

QoS differentiation opens new vulnerabilities to attacks referred to in the literature as ``QoS abuse'' \cite{Potrino2019, DeRango2020} or ``class hijacking'' \cite{Haywood2009, politis2016mac}.
In this area, researchers have also studied selfishness in forwarding multimedia streams \cite{Li2018}, nodes misreporting channel request parameters \cite{fang2016truthful}, and various cooperative forwarding strategies \cite{Kirchhof2020}.
Campus network designers have long foreseen that too much high-priority traffic may overwhelm the available bandwidth and/or switch capacity \cite{Szigeti2013}; in infrastructure-based networks under administrative supervision an obvious solution is to allow traffic marking with CoS/DSCP only at the network edge, subject to valid traffic contracts, rather than at users' premises. However, QoS abuse, exemplified by TRAs, is much harder to defend against in ad hoc networks, which lack a well-defined user-to-network interface.

TRAs are tied to the link layer: despite being launched at the Internet layer (through modifying DSCP), they exploit the underlying channel access prioritization.
Local-scope TRAs were considered in \cite{Konorski2014} and acknowledged as a threat to transmission opportunity sharing protocols \cite{Politis2018}. 
Multi-hop settings are also vulnerable \cite{Saxena2018}; \cite{Szott2014} discusses attacks in IEEE 802.11s mesh networks, \cite{Szott2018} -- in two-hop relay networks, while \cite{Szott2017} provides an overview of TRAs in ad hoc networks along with a discussion of attack detection and defense measures.
Additionally, \cite{Choi2020} studies a practical relaying scenario where users may execute TRAs.

From a detection viewpoint, TRAs are more challenging in multi-hop settings than in single-hop ones, because it is not always clear how local-scope manipulation of per-traffic class handling translates into end-to-end per-flow or per-packet performance -- link layer attacks such as TRAs consist in aggressive competition for a limited resource (the radio channel), bringing more benefit to the attacker and more harm to the honest nodes than do Internet layer attacks, where the benefit or harm is less pronounced. Single-hop settings are also easier to defend: when a traffic remapping attacker has been identified, it can be punished by neighbor honest nodes via responding in kind, e.g., increased transmission rate or jamming \cite{Patras2016}. In a multi-hop setting, however, punishment of TRAs may prove ineffective if known local-scope defense mechanisms are directly mimicked. Thus studies of selfish link layer attacks in multi-hop wireless networks leave many insights to be gained.

In this paper we are interested in MAWiN-oriented defense mechanisms justifiable by noncooperative game-theoretic considerations, i.e., rendering TRAs non-beneficial for attackers in terms of perceived QoS. This requires simple performance models of MAWiNs under TRAs that yield closed-form solutions and thus handy payoff functions for arising games. Few such models are known, none of them able to capture on a macroscopic level the complex interplay of channel access queuing and contention, EDCA prioritization, node mobility, and intra-flow competition due to multi-hop forwarding in the presence of hidden nodes (where packet transmissions from one node compete with those from up- and downstream nodes one or two hops away). Existing models are usually limited to chain topologies \cite{Shimoyamada2015, Sanada2018} or tied to specific analytical models of high complexity \cite{Rezaei2018}; they do not consider traffic differentiation.

\section{Network Model}
\label{sec:model}
In this section we formalize the network and TRA description, and develop an end-to-end performance model to quantify TRA motivation and impact.
A summary of the notation used in the paper is presented in Table~\ref{t_notation}.

\begin{table}[t]
\centering
\caption{Summary of Notation Used}
\begin{tabular}{@{}lp{0.75\columnwidth}@{}}
\toprule
Symbol & Definition \\ \midrule
$ac$ & An e2e-flow's intrinsic AC \\
$A$ & Set of attackers \\
$cost$	& Nodal cost metric \\
$CH_i(r,ac)$ & Set of h-flows competing with outgoing h-flow $(i, r, hac_i(r, ac))$ \\
$s_r,d_r$      & Source and destination nodes of route $r$ \\
$\Delta$	& Set of in-distress nodes \\
$\Delta^*$	& Set of in-exposure nodes \\
$E_{R_F}(M)$ & Set of nodes whose source traffic is forwarded by nodes in $M$   \\
$E_{R^*_F}(M)$ & Set of nodes forward-reliant on nodes in $M$   \\
$fcost$	& e2e-flow cost metric \\
$F$		& Set of e2e-flows in the network \\ 
$F^*$		& Set of survivable e2e-flows in the network \\ 
$G(k)$	& Set of in-game nodes in stage $k$ \\
$\Gamma$	& Set of best-reply nodes \\
$\mathbf{h}_i$	& Node $i$'s history of membership in $A(k-1)$ and $A(k)$ \\
$H$ 	& Set of recognizable h-flows \\
$I$	& Set of ill-behaved nodes \\
$(j, r, hac)$ & An h-flow of e2e-flow $(r, ac)$ incoming from node $j$\\
$k$ 	& Stage of the TRA game \\
$mang()$ & Packet mangling function (CoS-to-AC mapping) \\
$map$ & Function determining  ACs of outgoing h-flows  \\
$OH_i$	& Set of outgoing h-flows at node $i$ \\
$pred_{r,i}$ & Predecessor (previous-hop node) to node $i$ on route $r$ \\
$P_{r,i}$ & Set of nodes that precede or coincide with node $i$ on route $r$ \\
$\sigma$	& An action selection rule in the TRA game \\
$succ_{r,i}$ & Successor (next-hop node) to node $i$ on route $r$ \\
$r$, $||r||$      & An end-to-end route and its hop-length \\
$(r, ac)$	& An e2e-flow of intrinsic AC $ac$ following route $r$  \\
$rank_i(r, ac)$ & Performance metric for e2e-flow $(r,ac)$ at node $i$ \\
$R$      & Set of all end-to-end routes in the network \\
$R_F$	& Forwarding relationship \\
$R_F^*$	& Forward-reliance relationship \\
$T = \langle N,L \rangle$      & Network topology graph (set of nodes, set of links representing node hearability)           \\
\bottomrule
\end{tabular}
\label{t_notation}
\end{table}

\subsection{Topology, Routes, and Flows}
\label{sec:topology}
A static MAWiN topology is represented by a directed graph $T=\langle N,L \rangle$, where $N$ is the set of nodes, $L \subset N \times N$, and $(i, j) \in L$ iff $i \neq j$ and $j$ is in the hearability range of $i$. Let $N^*$ be the set of all directed acyclic routes in $T$ and $R \subseteq N^*$ be the set of end-to-end routes in $T$ as determined by the routing algorithm in use. Each route $r \in R$ is represented as a sequence of nodes $r=(i_1, \dots, i_m)$ such that $i_1, \dots, i_m$ are all distinct and $(i_{m'}, i_{m'+1}) \in L$ for all $m'=1, \cdots, m-1 $; $i_1$ and $i_m$ are the source and destination nodes of $r$, denoted $s_r$ and $d_r$, $i_2, \dots, i_{m-1}$ are the transit nodes, and $\lVert r \rVert = m - 1$ is the hop length of $r$.
We write $i \in r$ if $r$ involves node $i$; for $i, j \in r$ write $i <_r j$ ($i \leq_r j)$ if $i$ precedes (precedes or coincides with) $j$ on $r$. For $i \in r \setminus \{d_r\}$ denote by $succ_{r,i}$ the immediate successor of $i$ on $r$, and for $i \in r \setminus \{s_r\}$ define $pred_{r,i}$ as the immediate predecessor of $i$ on $r$; for uniformity, $pred_{r,s_r}$ is defined as $s_r$.
(In the parlance of wireless multi-hop networks, a node $i'$ is said to be \emph{hidden} from $i$ if $(i, i') \notin L$ and $\exists j \in N,r,r' \in R$ : $j=succ_{r,i}=succ_{r',i'}$.) 
Denote by $P_{r,i }= \{j | j \leq_r i\}$
the set of nodes that precede or coincide with $i$ on $r$.

We model network traffic as composed of \emph{end-to-end (e2e-) flows}, each of which is a collection of packets of the same CoS $\in \{\text{EF},\text{BE}\}$ and moving along the same route.
The corresponding link layer frames are handled by EDCA according to assigned ACs, contained in the AC fields of their headers.
For ease of presentation we restrict the used ACs to \emph{VO} (real-time traffic such as voice/video) and \emph{BE} (best-effort traffic), with \emph{VO} having (statistical) priority over \emph{BE} at the link layer. Since packet mangling amounts to a CoS-to-AC mapping, we define a function $mang: \{\text{EF},\text{BE}\} \to \{VO,BE\}$ such that $mang(\text{EF}) = VO$ and $mang(\text{BE}) = BE$.
An e2e-flow of a given CoS is represented as $(r, ac)$, where $r \in R$ is its route and $ac = mang(\text{CoS})$ is its intrinsic AC as returned at $s_r$.  Let $F \subseteq R \times \{VO,BE\}$ be the (quasi-static) set of e2e-flows offered by MAWiN users. Presumably, only
nodes generating their own source traffic are interested in staying connected, therefore we assume that at least one e2e-flow is offered at each node, i.e.,
\begin{equation}
\label{flowsources}
\{s_r | (r, ac) \in F\} = N. 
\end{equation}

For further analysis, it is necessary to formally state the fact that a source of a traffic flow is reliant on a set of nodes for forwarding its packets.
\begin{definition}
Let $R_F \subseteq N \times N$ be a binary relationship such that $(i, j) \in R_F$ iff $j$ forwards $i$'s source traffic, i.e., $\exists(r, ac) \in F: i = s_r \land j \in r \setminus \{d_r\}$ (in particular, $(s_r, s_r) \in R_F$). The transitive closure of $R_F$, denoted $R_{F}^*$, will be referred to as \emph{forward-reliance}, i.e., $(i,j) \in R^*_F$  iff there exists a sequence of nodes $i_1,i_2,...,i_k$ with $i_1=i$ and $i_k=j$ such that $i_{l+1}$ forwards $i_l$'s source traffic, $l=1,\dots,k-1$.
\label{def:reach}
\end{definition}
Assume that a removal from the network of a forwarding node on $r$ causes $s_r$ to be removed as well. Then $(i,j) \in R^*_F$ expresses node $i$'s forward-reliance on node $j$ in that a removal of $j$ ultimately causes a removal of $i$.
For $M \subseteq N$, let $E_{R_F}(M)=\{i \in N | \exists j \in M:(i,j)\in R_F\}$; the set of nodes forward-reliant on nodes in $M$ is therefore $E_{R_{F}^*}(M)$. We have $M \subseteq E_{R_F}(M) \subseteq E_{R_{F}^*}(M)$, the first inclusion due to (\ref{flowsources}). 

We use the notion of \emph{hop (h-) flows} as the granulation level at which incoming traffic is recognized at a next-hop node. %
Packets of e2e-flow $(r, ac)$ forwarded by $j=pred_{r,i}$, whose AC fields contain $hac \in \{VO,BE\}$, are recognized at node $i \in r$ as an h-flow $(j, r, hac)$. (Possibly $hac \neq ac$, because AC fields can be modified hop-by-hop.) 
By convention, let e2e-flow $(r, ac)$ be recognized at $s_r$ as h-flow $(s_r,r,ac)$. For example, if node 3 in Fig.~\ref{fig:topology} changes the AC fields of incoming packets then e2e-flow \#1, designated as (1, $VO$), is recognized as (1, $r_1$, $VO$) at node 1, (1, $r_1$, $VO$) at node 3, (3, $r_1$, $BE$) at node 4 etc., where $r_1=(1,3,4,5,8,10)$. Let $H \subseteq N \times R \times \{VO,BE\}$ be the set of recognizable h-flows.

Autonomous operation of node $i$ is expressed as a function $map_i: H \rightarrow \{VO,BE\}$ according to which it sets ACs of h-flows. For an incoming h-flow $(j, r, hac)$, where $j = pred_{r,i}$ and $i \in r \setminus \{s_r, d_r\}$, the new AC field forwarded by $i$ further along $r$ is given by $map_i(j, r, hac)$.

\subsection{Attack Model}
\label{sec:attacks}
We consider an attacker to be a selfish node which aims to receive a higher QoS level for its source traffic by performing a TRA, i.e., changing the traffic class of incoming transit or source flows. The Internet layer's packet mangling functionality is used as explained in Section~\ref{sec:introduction} (Fig.~\ref{fig:tra-stack}) to interfere with the default CoS-to-AC mapping and modify the IP header fields of incoming packets, which are transmitted further along their path with a different MAC-layer priority. Except for the path's destination node, any node can potentially become an attacker. We assume that:
\begin{itemize}
\item a MAWiN node is capable of assessing received QoS,
\item in terms of received QoS, a TRA can be beneficial to the attacker and harmful to other nodes,
\item a TRA can be performed at no expense and at no risk of detection or administrative punishment, and
\item a subset of nodes are selfish and ready to become attackers.
\end{itemize}
A TRA can be formally described as follows.

\begin{definition}
A \emph{traffic remapping attack} (TRA) that a node $i \in r \setminus \{d_r\}$ launches upon an incoming h-flow $(j, r, hac)$, where $j = pred_{r,i}$, consists in changing its AC, i.e., configuring $map_i(j, r, hac) \neq hac$.
\label{def:intro-tra}
\end{definition}

Such a definition captures the fact that the setting of AC fields under a
TRA is both protocol compliant (the use of $map_j$ is feasible) and ill-willed (inconsistent with $mang(\cdot)$).

We consider the behavior of a node which does not perform a TRA to be honest, whereas attackers can either upgrade or downgrade an incoming h-flow's AC. We formally define these behaviors as follows.

\begin{definition}
The behavior of node $i$ with respect to h-flow $(j, r, hac)$ can be classified as (i) \emph{honest}, if the AC remains unchanged (i.e., $map_i(j, r, hac) = hac$), (ii) \emph{upgrading TRA} (TRA$^+$), if $hac = BE$ and $map_i(j, r, BE) = VO$, or (iii) \emph{downgrading TRA} (TRA$^-$) if $hac = VO$ and  $map_i(j, r, VO) = BE$. Nodes that exhibit behavior (ii) or (iii) will be called \emph{attackers}.
\label{def:tra}
\end{definition}

We adopt a simple model of an attacker: it will launch a TRA on all its source and transit flows provided that the former can be upgraded and the latter downgraded.

\begin{definition}
An attacker is called \emph{plausible} if it never downgrades its own source traffic or upgrades transit traffic, i.e., $map_i(j, r, hac) = hac$ if $(hac = VO$ and $i = s_r)$ or $(hac = BE$ and $i \neq s_r)$, and \emph{opportunistic} if it launches a TRA$^+$ or a TRA$^-$ upon all h-flows it recognizes, subject to the plausibility constraints.
\label{def:plausible}
\end{definition}

In our model, each attacker is assumed to be plausible opportunistic.
Let $A\subseteq N$ denote the set of attackers.
Given $A$, the new AC field $hac_i(r, ac)$ is derived as:

\begin{equation}
\begin{aligned}
\label{hacBEVO}
&hac_i(r,BE)=\left\{
\begin{aligned}
& VO,\quad s_r\in A\wedge P_{r,i} \setminus \{s_r\} \cap A=\varnothing, \\
& BE,\quad \text{otherwise},
\end{aligned} \right. \\
&hac_i(r,VO)=\left\{
\begin{aligned}
& VO,\quad P_{r,i} \setminus \{s_r\} \cap A=\varnothing, \\
& BE,\quad \text{otherwise}.
\end{aligned} \right.
\end{aligned}
\end{equation}
That is, $hac_i(r, BE) = VO$ if a TRA$^+$ at $s_r$ and no TRA$^-$ have been launched by the time the flow's packets reach $i$, and $hac_i(r, BE) = BE$ if no TRA or both a TRA$^+$ and a TRA$^-$ have been launched.
Similarly, $hac_i(r, VO) = VO$ if no TRA$^-$ has been launched at nodes other than $s_r$, and $hac_i(r, VO) = BE$ if a TRA$^-$ has been launched\footnote{This type of downgrading attack is detectable at $d_r$ by comparing the $hac$ of the incoming h-flow with the flow's intrinsic AC (e.g., sent from $s_r$ to $d_r$ as encrypted metadata). While feasible, such detection is highly troublesome \cite{Szott2017} and unable to pinpoint the attacker.}.

\begin{figure}[tbp]
\centering

\includegraphics[width=\columnwidth]
{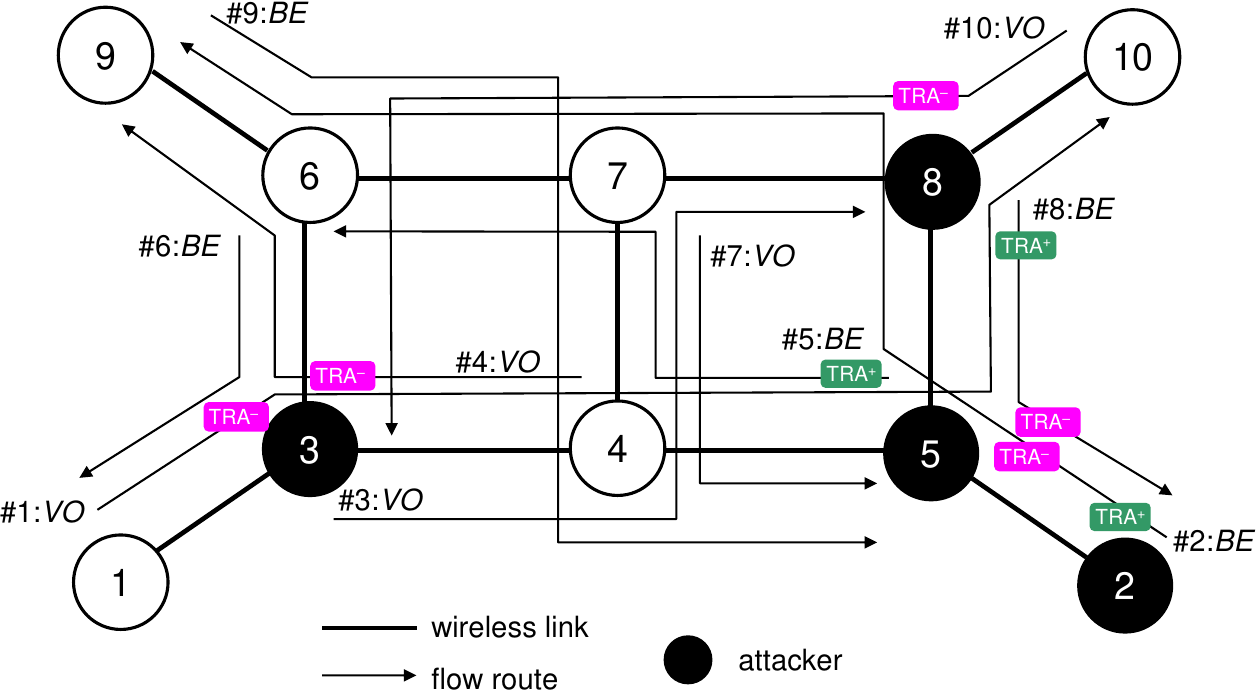}%
\caption{Example MAWiN topology, e2e-flow routes, and experienced TRAs; node $i$ is the source of e2e-flow \#$i$ with intrinsic AC indicated.}
\label{fig:topology}
\end{figure}

For a 10-node MAWiN with $|N| = |F| = 10$, route hop lengths varying from $\lVert r \rVert_{\text{min}} = 2$ to $\lVert r \rVert_{\text{max}} = 5$, and $A = \{2,3,5,8\}$, Fig.~\ref{fig:topology} displays
TRAs experienced by each e2e-flow. One notes in particular that:

\begin{itemize}
\item e2e-flow \#3 with $ac=VO$ has an attacker source which, however, does not launch a TRA$^{-}$ due to the plausibility constraints,

\item likewise, e2e-flows \#3, \#7, and \#9 have an attacker destination which behaves honestly due to the plausibility constraints,

\item e2e-flow \#6 is not attacked at an attacker transit node 3, which could only have launched a TRA$^{+}$, %

\item e2e-flow \#1 with $ac = VO$ encounters three attacker transit nodes, of which the first launches a TRA$^{-}$, hence the second and third no longer have to,

\item e2e-flows \#2 and \#8 experience a TRA$^{+}$ at their source nodes and a TRA$^{-}$ at node 5; this is the maximum number of attacks an e2e-flow can experience. 
\end{itemize}

\subsection{End-to-End Performance Model}
\label{sec:performance}

Evaluation of the impact of and countermeasures against TRAs requires an analytical performance model of a MAWiN in the presence of TRAs that can deal with arbitrary topologies $T$ and flow sets $F$, and thus obviate the need for time consuming and setting specific full-stack simulations. Motivated by the deficit of such models in existing literature, cf. Section~\ref{sec:soa}, and building on the models of Section~\ref{sec:topology} and Section~\ref{sec:attacks}, we have developed an approximate rank-based model, described below. Each h-flow is assigned a rank depending on the number and priorities of h-flows it has to compete with for channel access; the collection of ranks of all h-flows constituting a given e2e-flow is then translated into an informative end-to-end QoS metric.
At node $i$, the set of outgoing h-flows is
\begin{equation}
\label{OH}
OH_i = \{(i, r, hac_i(r, ac)) | (r, ac) \in F \wedge i \in r \setminus \{d_r\}\}.
\end{equation}
For an outgoing h-flow $(i, r, hac_i(r, ac))$, the set $CH_i(r,ac)$ of competing h-flows consists of: (a) other outgoing h-flows at $i$, which compete via the local transmission queue, (b) outgoing h-flows at nodes in the hearability range of $i$, which compete via CSMA/CA, and (c) outgoing h-flows at nodes hidden from $i$, which compete via exclusive-OR reception at $succ_{r,i}$:
\begin{multline}
CH_i(r,ac) = OH_i \setminus \{(i,r,hac_i(r,ac))\} \\
 \cup \bigcup\nolimits_{j:(j,i)\in L}OH_j
\cup \bigcup\nolimits_{j:(j,succ_{r,i})\in L \land (j,i) \not\in L}OH_j.
\end{multline} 
For the above outgoing h-flow, the pair $[hac, CH]_i(r, ac)$ determines per-hop performance at node $i$, where we use a succinct notation $[a, b]_i(x)$ instead of $[a_i(x), b_i(x)]$. We propose a per-hop performance metric $rank_i(r, ac)$ reflecting that an h-flow is better off at a node if it is \emph{VO}, and competes with fewer and preferably \emph{BE} h-flows. Accordingly, the metric should rank $[hac,vo,be]_i(r, ac)$ vectors, where $vo_i(r, ac)$ and $be_i(r, ac)$ represent the number of \emph{VO} and \emph{BE} h-flows in $CH_i(r, ac)$.

To evaluate $rank(\cdot)$, we used the Markovian model of EDCA \cite{szott2010ieee} to calculate the normalized per-hop saturation throughput $S_i(r, ac)$ of e2e-flow $(r,ac)$ at node $i$ for various $hac=hac_i(r,ac) \in \{VO,BE\}$, $vo=vo_i(r, ac) \in \{0,\dots,10\}$, and $be=be_i(r, ac) \in \{0,\dots,10\}$. For the resulting 14,520 pairs of $(hac, vo, be)$ vectors, $rank(\cdot)$ represents a good fit if
\begin{multline}
\label{implic}
rank_i(r, ac)|_{hac,vo,be} \le rank_i(r, ac)|_{hac',vo',be'}
\\
\text{iff}~S_i(r, ac)|_{hac,vo,be} \ge S_i(r, ac)|_{hac',vo',be'}
\end{multline}
holds for a high percentage of the pairs. (Obviously, a small rank is desirable.) A heuristic metric is
\begin{multline}
\label{rank}
rank_i(r, ac)=\mathbbm{1}_{hac=BE} \cdot \alpha \cdot (vo+\mathbbm{1}_{vo>1 \lor be>2})
\\
+ \beta \cdot (vo+\mathbbm{1}_{hac=BE})+be,
\end{multline}
where $\mathbbm{1}_{(\cdot)}$ is the indicator function, and the best fit (99.13\% of the pairs) occurs at $\alpha =40$ and $\beta =10$. The preferences of h-flows are reflected: $vo$ has more impact upon $rank_i(r, ac)$ than does $be$ (because $\beta > 1$), and there is distinct separation between $hac=VO$ and $hac=BE$ (because $\alpha \gg \beta$). 

For any $A \subseteq N$, $rank(\cdot)$ induces a heuristic e2e-flow cost metric we call $fcost$, additive for \emph{VO} traffic delay and bottleneck-type for \emph{BE} traffic throughput, defined as:
\begin{equation}
\label{fcost}
fcost_{(r,ac)}(A)=
\left\{
\begin{aligned}
& \frac{\sum_{i\in r\setminus\{d_r\}} rank_i(hac,r,ac)}{\lVert r \rVert - 1}, &ac=VO, \\
& \max_{i\in r\setminus\{d_r\}}rank_i(hac,r,ac), &ac=BE,
\end{aligned} \right. 
\end{equation}
where $hac$ is given by (\ref{hacBEVO}) and the notation $fcost_{(r,ac)}(A)$ is meaningful, because $hac$ depends on $A$. From (\ref{fcost}), a nodal cost metric $cost$ can be derived as a weighted sum
\begin{equation}
\label{cost}
cost_i(A)=\sum_{(r,ac) \in F: s_r=i}\gamma_{r,i} \cdot fcost_{(r,ac)}(A), 
\end{equation}
where $\gamma_{r,i} \ge 0$ and $\sum_r\gamma_{r,i}=1$.
The status of an attacker (honest) node whose cost relative to the $A = \varnothing$ case has increased is \emph{lose} (\emph{mind}), otherwise it is \emph{don't lose} (\emph{don't mind}).

\begin{figure}[t]
\centering
\includegraphics[width=\columnwidth]
{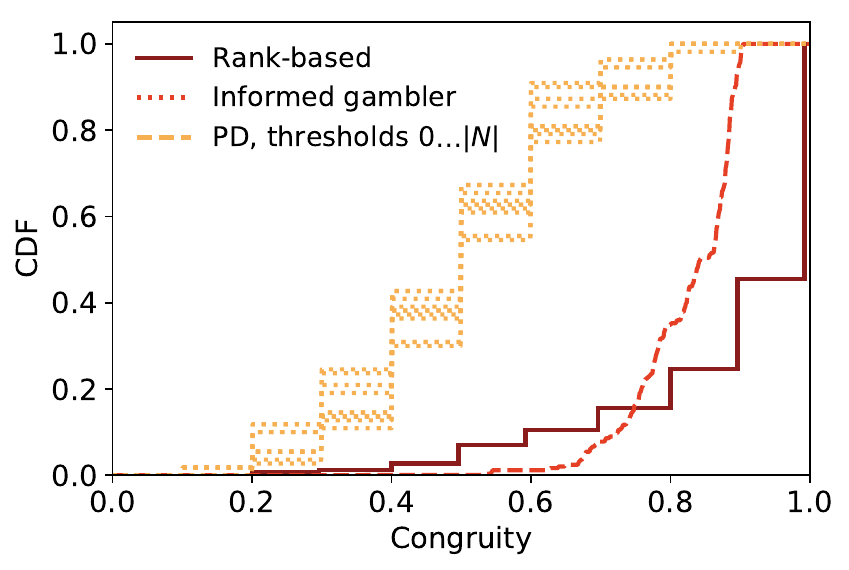}%
\caption{Congruity between simulations and various analytical models.}
\label{fig:validation}
\end{figure}

To validate the rank-based model, 
we implemented the network topology and flow set of Fig.~\ref{fig:topology}
in the ns-3 simulator, assuming error-free radio channels, static routing, and constant bit-rate saturation-level UDP traffic of 1500 B packets.
Each simulation run lasted 200~s with an additional 50~s warm-up time and was repeated five times. Nodes were classified as \emph{mind} or \emph{lose} if the throughput of their BE flows dropped by 5\% or more, or if the per-hop delay of their VO flows increased by more than 20 ms and exceeded 100 ms.   
We assessed \emph{congruity}, defined as the proportion of nodes whose status (\emph{mind}, \emph{lose}, \emph{don't mind}, or \emph{don't lose}) upon TRAs launched by a random attacker set agrees between the simulation and the rank-based model.
Fig.~\ref{fig:validation} presents the cumulative distribution function (CDF) of congruity obtained after simulating all the 256 possible attacker sets\footnote{Nodes 1 and 10 cannot launch a TRA$^+$
or a TRA$^-$
due to the plausibility constraints of Definition~\ref{def:plausible}.}, producing a mean congruity of 0.89.

For comparison consider a heuristic inspired by the $|N|$-person Prisoners' Dilemma (PD) game \cite{Straffin1993}:
if the number of attackers exceeds (does not exceed) a certain threshold then all the attacker nodes' status is guessed as \emph{lose} (\emph{don't lose}) and the honest nodes' as \emph{don't mind} (\emph{mind}). 
The corresponding CDFs depicted in Fig.~\ref{fig:validation} for the threshold varying from $0$ to $|N|$ produce mean values between 0.49 and 0.54, not far from a fair coin toss. 
As another baseline, an unrealistic ``informed gambler'', who knew an attacker (honest) node's statistical chance of acquiring a \emph{lose} (\emph{mind}) status under a random attacker set,
might guess the node's status for a given $A$ by tossing an appropriately biased coin. Congruity would then be
measured by the expected number of guesses that match the simulation. The corresponding CDF
depicted in Fig.~\ref{fig:validation} produces a mean value of 0.82, inferior to our model's.
We conclude that the rank-based model is a reasonably good predictor of the impact of TRAs in MAWiNs with saturation-level traffic.

It will be convenient to identify nodes directly impacted by a given attacker set.
\begin{definition}
A node whose status is \emph{lose} or \emph{mind} is said to be \emph{in distress}. Let $\Delta(A)$ be the set of such nodes in the presence of the attacker set $A \subseteq N$.
\label{auxil1}
\end{definition}
Hence, the set of in-distress nodes contains nodes whose costs have increased in comparison to the $A = \varnothing$ case: $\Delta(A) = \{i \in N | cost_i(A) > cost_i(\varnothing)\}$. Note that %
$\Delta(A) = \varnothing \neq A$ is possible, e.g., if $A = \{1,10\}$ in Fig.~\ref{fig:topology}.

\subsection{Attack Incentives and Impact}
\label{sec:motivation}

In realistic MAWiN settings, TRAs pose a threat whose credibility (i.e., incentives to launch) and seriousness (i.e., harmful impact upon honest nodes) we now quantify. We ask if, regardless of the currently ongoing TRAs, an honest node turning attacker perceives a QoS improvement and causes some other nodes to perceive a QoS degradation. Neither of these effects is certain, as it depends on a node's position in the network topology. Referring to Fig.~\ref{fig:topology}, suppose that $A= \varnothing$ and node 3 turns attacker. Since flow \#3 is VO, the attack amounts to a $\text{TRA}^-$ upon flows \#1 and \#4. While these two flows suffer, for all the remaining flows the contention softens and their QoS improves. Hence, the TRA is harmless for other nodes. Consider an alternative scenario where it is node 5 that turns attacker. Its source traffic (flow \#5) now enjoys elevated priority when forwarded at nodes 5, 4, and 7, but  experiences increased contention from itself at node 5 (being forwarded by node 4 and via  exclusive-OR reception from node 5 at node 4) and at node 4 (being forwarded by nodes 5 and 7), the likely net effect of which is a QoS degradation.

\begin{figure*}[t]
\centering
\subfloat{\includegraphics[width=0.5\textwidth]{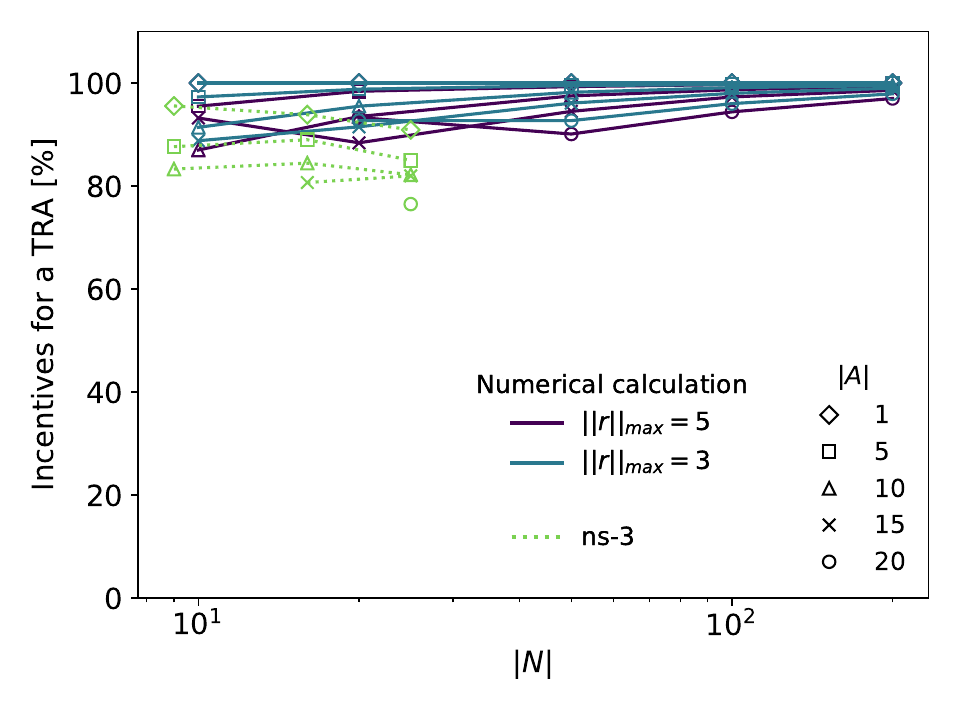}%
\label{fig-motivation-perA-incentives}}
\subfloat{\includegraphics[width=0.5\textwidth]{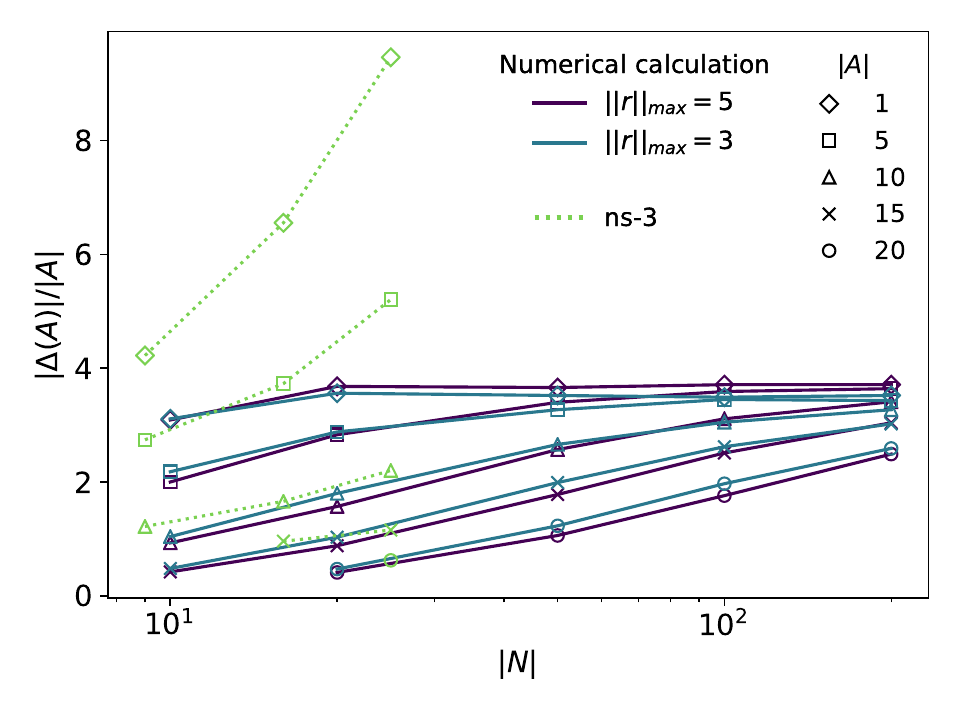}%
\label{fig-motivation-perA-indistress}}
\vspace{-0.6cm}
\caption{
Incentives to launch a TRA (\textit{left}), harmful impact (``direct victims'') of a TRA (\textit{right}); $\lVert r \rVert_{\text{min}} = 1$.}
\label{fig:scaling1}
\end{figure*}

To investigate the above effects and their scaling with the network size $|N|$, both numerical calculation using the cost metric (\ref{cost}) and ns-3 simulations were conducted. In the numerical calculation we assumed $|N| \le 200$, uniformly distributed route hop lengths with $\lVert r \rVert_{\text{min}} =$ 1, and  $\lVert r \rVert_{\text{max}} =$ 3 or 5, and $|A|$ = 1, 2, 5, 10, or 20. For a fixed parameter configuration, the results were averaged over 10,000 instances of random network topologies, e2e-flow routes, and attacker sets. The transit nodes for a route were chosen at random in geographical proximity to the source node. We examined the scaling with $|N|$ of:
\begin{itemize}
    \item the incentives to launch a TRA, measured as the percentage of attacker nodes whose cost metric did not worsen when turning attacker, i.e., nodes $i \in A$ satisfying $cost_i(A) \le cost_i(A \setminus \{i\})$, and
    \item the harmful impact of a TRA launched in the presence of a number of ongoing TRAs, measured as $|\Delta(A)|/|A|$, i.e., the average \emph{in-distress} nodes per attacker node.
\end{itemize}

The results depicted in Fig.~\ref{fig:scaling1} indicate that the threat of TRAs is not limited to small-size networks. The incentives for a TRA remain 100\% for $|A| = 1$ and decrease with $|A|$, but slightly increase with $|N|$ on account of more dispersed attacker nodes. They also slightly increase with the route hop length. A similar scaling, insensitive to route hop lengths, is visible for the harmful impact of a TRA.

For the ns-3 simulations we assumed square grid topologies with $|N| =$ 9, 16, and 25. 
Each node was the source of an e2e-flow following a minimum-hop route to a randomly chosen destination, and half of the flows were $VO$ (other settings are described in Section \ref{sec:simulations}). As before, we used end-to-end throughput and delay as QoS metrics for \emph{BE} and \emph{VO} flows, respectively, and the same rules of deciding node status. The results are marked in Fig.~\ref{fig:scaling1} with dotted lines. 
The	incentives for TRAs are somewhat lower now, reflecting the fact that nodes located at the edge of the square grid have no transit traffic to downgrade.
Meanwhile, in-distress nodes are more numerous, reflecting inter-flow competition effects unaccounted for by the cost metric. 
Nonetheless, these results confirm that the incentives for and harmful impact of TRAs are significant for large $|N|$.

\section{TRA Game Description}
\label{sec:game}

A selfish node performs a TRA whenever this improves its cost, anticipating similar conduct of other selfish nodes. This gives rise to a noncooperative \emph{TRA game}, whose one-shot and multistage variants we now describe. We propose to mitigate TRAs by introducing a distributed exposure signaling mechanism called DISTRESS, under which ``good'' multistage strategies lead to few nodes performing TRAs when the game terminates.

\subsection{One-Shot TRA Game}
\label{sec:one-shot-game}
In the noncooperative one-shot TRA game, the nodes are players, $map_i \in \{TRA, honest\}$ is node $i$'s action, and \emph{cost} is the (negative) payoff function (i.e., small costs are pursued). 
A given action profile $(map_i, i \in N)$ is equivalent of the set $A \subseteq N$ of plausible opportunistic attackers (nodes launching TRA).
Using a $\langle \text{players, action space, payoffs}\rangle$ representation, the game is defined as
\begin{equation}
\label{gamedef}
\langle N, 2^{N}, cost: N \times 2^{N} \rightarrow \mathbf{R}^+\rangle.
\end{equation}
Some interesting action profiles are: $\varnothing$ (all-honest), and $N$ (all-TRA).
In the latter, any e2e-flow $(r, BE)$ experiences a TRA$^+$ at $s_r$, and any e2e-flow $(r, VO)$ experiences a TRA$^-$ at the first encountered node in $r \setminus \{s_r, d_r\}$.

Contrary to the intuition
that it is always beneficial to upgrade source traffic 
and downgrade competing transit traffic,
the TRA game is not an $|N|$-person PD.
Specifically, due to the complexities of mechanisms determining MAWiN performance, the $TRA$ action does not dominate $honest$, %
nor is it necessarily harmful to honest nodes. %
Moreover, $A=\varnothing$ need not be Pareto superior to $A=N$; in fact, for some traffic patterns,
the reverse is true \cite{Konorski2017a}.

The following definition identifies nodes indirectly impacted by a given attacker set.
\begin{definition}
A node forward-reliant on a node in distress (cf. Definitions~\ref{def:reach} and \ref{auxil1}) is said to be \emph{in exposure}. Let $\Delta^*(A)$ be the set of such nodes in the presence of the attacker set $A \subseteq N$.
\label{auxil2}
\end{definition}
Hence, the set of in-exposure nodes contains nodes which are forward-reliant on nodes currently in distress: $\Delta^*(A) = E_{R_{F}^*}(\Delta(A))$. The rationale behind Definition~\ref{auxil2} is the following. 
Nodes in distress perceive unsatisfactory QoS and so lose incentives to continue packet forwarding services. Thus they pose a credible threat of imminent \textit{service suspension}, to which exposed are they themselves along with the set of source nodes whose traffic they forward. In the case of service suspension by a node in distress, each of these source nodes registers an infinite e2e-flow cost (\ref{cost}), and also loses incentives to continue packet forwarding services. By recursion, similarly exposed are all nodes forward-reliant on nodes in $\Delta(A)$, i.e., $\Delta^*(A)$.

TRAs can be mitigated by leveraging node exposure in a way not unlike an immune response is triggered by a foreign toxin. Namely, even a small attacker set creates a ripple effect across the network, causing exposure in a much larger set of nodes than those in distress. Instead of suspending service, nodes in exposure start playing the TRA game, occasionally selecting $TRA$ rather than $honest$, which may cause exposure in the initial attackers. If exposure, i.e., the threat of imminent service suspension, is reflected in the game payoffs as a large enough (say infinite) $cost$, such play brings most or all of the attackers back to honesty, which the nodes in distress alone may be too few to achieve. A rigorous argument is given in Section~ \ref{sec:multistage-strategy}.

\begin{figure}[t]
\centering
\includegraphics[width=\columnwidth]{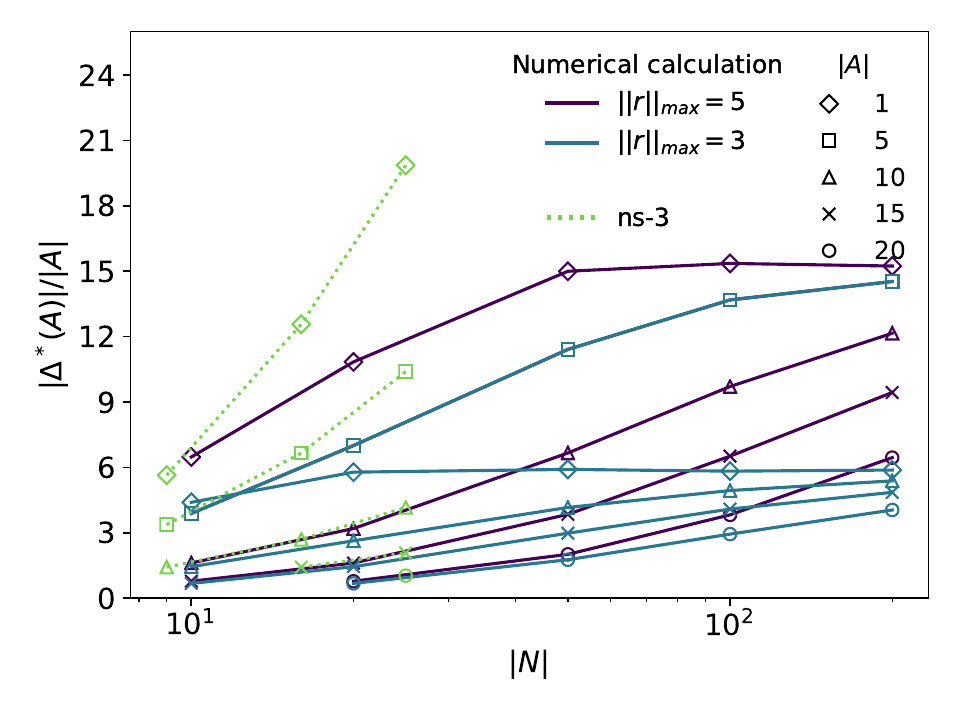}%
\caption{
Size of the ``immune response''; $\lVert r \rVert_{\text{min}} = 1$.
}
\label{fig:scaling2}
\end{figure}

For various $|N|$, we examined the size of the ``immune response'' triggered by a TRA, measured as $|\Delta^*(A)|/|A|$, the average in-exposure nodes per attacker node. Using cost metric-based calculation we verified that it remains nontrivial and does not distinctly decrease at least up to $|N| = 200$ (especially under longer routes), cf. Fig.~\ref{fig:scaling2}. In the ns-3 simulations of square grid topologies up to $|N| = 25$, $\Delta^*(A)$ was inferred by keeping track of the source nodes of flows currently forwarded by each node. The results (marked with dotted lines) show a nontrivial size of the ``immune response'', which even scales with $|N|$. 

To incorporate exposure into the game payoffs we redefine nodal costs (\ref{cost}) and the TRA game (\ref{gamedef}) as
\begin{equation}
cost'_i(A)=\left\{
\begin{aligned}
&cost_i(A),&i\notin \Delta^*(A), \\ 
&\infty, &i\in \Delta^*(A),
\end{aligned}
\right.
\label{eq:cost-redefined}
\end{equation}
\begin{equation}
\label{gamemod}
\langle N, 2^{N}, cost': N \times 2^{N} \rightarrow \mathbf{R}^+\rangle.
\end{equation}

\subsection{Multistage Play under the DISTRESS Mechanism}
\label{sec:multistage-game}
Mitigation of TRAs using the above approach requires that nodes signal exposure to one another and can toggle between $TRA$ and $honest$ in response to other nodes' play, as modeled by a multistage game. Let the TRA game (\ref{gamedef}) be played in stages $k = 1, 2, {\dots}$, and
let $A(k) \subseteq N$ be the set of attackers in stage $k$.
Each stage $k$ is assumed long enough for each node $i$ to produce an accurate estimate of $cost_i(A(k))$ and signal exposure
throughout the network if needed, hence to also determine $cost'_i(A(k))$.

Assume that there are no attackers prior to stage 1, i.e., $A(0) = \varnothing$.
In stage 1, a set $I \subset N$ of \textit{ill-behaved} (selfish) nodes spontaneously select $TRA$, i.e., $A(1) = I$; the other nodes are further called \textit{well-behaved}.
The stage-1 TRAs may bring about distress in some (possibly also ill-behaved) nodes, i.e., induce the set $\Delta(A(1)) = \Delta(I)$.
At the end of a generic stage $k - 1$, each node $i$ estimates its current cost metric and if it finds itself in distress, i.e., $i \in \Delta(A(k - 1))$, then marks
itself as in-exposure and sends a DISTRESS$(i)$ flag to all nodes whose source traffic it forwards; that is, copies of the flag are sent to all nodes of $E_{R_F}(\{i\})$, including node $i$ itself. (The flag is also timestamped to avoid confusion with exposure signaling in another stage.) 
Having received DISTRESS$(j)$ and checked that $(i,j) \in R_F$, node $i$ ignores the flag if it is already in exposure, otherwise marks
itself as in-exposure and sends DISTRESS$(i)$ to $E_{R_F}(\{i\})$ as above. It is easy to see that if $M$ is the set of nodes so far marked as in-exposure then the process stops when $E_{R_F}(M) = M$. Thus, assuming that each flag is issued and reliably delivered to the intended recipient within one stage, at the end of stage $k$ we have $M=\Delta^*(A(k - 1))$. 
This mechanism, specified as Algorithm~\ref{algorithm}, will be termed DISTRESS (DIstributed Signaling of TRaffic Exposure to Service Suspension).

\begin{algorithm}[t]
\small
\SetAlgoLined
\algorithmfootnote{For ease of specification, a DISTRESS flag is assumed to be sent only once and forwarded reliably to the intended recipient.}
\KwData{$A(k-1)$ -- set of attackers in stage $k-1$}
\KwResult{$\Delta^*(A(k-1))$ -- set of nodes in exposure at the end of stage $k$}

$M \leftarrow \varnothing$ \tcp*[l]{no node marked as in-exposure}

\Repeat{$E_{R_F}(M)=M$}{
    \For{$i \in N \setminus M$}{
        \If{$(cost_i(A)>cost_i(\varnothing)) \lor \newline ((\emph{DISTRESS}$(j)$ \emph{ received}) \land ((i,j) \in R_F))$}{
            $M \leftarrow M \cup \{i\}$ \tcp*[l]{node $i$ marked as\ 
            in-exposure} \
            send DISTRESS$(i)$  to  $E_{R_F}(\{i\}$)
            
         } 
    }    
}
$\Delta^*(A(k-1)) \leftarrow M$

\caption{DISTRESS mechanism}
\label{algorithm}
\end{algorithm}

It is vital that the threat signified by signaled exposure be credible. Therefore a received DISTRESS$(j)$ flag, where $j \in r \setminus \{s_r\}$, should imply to $s_r$ that $j$ is indeed a transit node on $r$ and not an off-route one that does not pose a threat. This is granted if a source-routing protocol such as DSR \cite{rfc4728} is employed, in which $r$ is known to $s_r$. Otherwise, DISTRESS$(j)$ can be appended to forwarded packets and at $d_r$ returned to $s_r$ through a (trusted) end-to-end feedback connection. 
\begin{figure}[tbp]
\centering
\includegraphics[width=\columnwidth]
{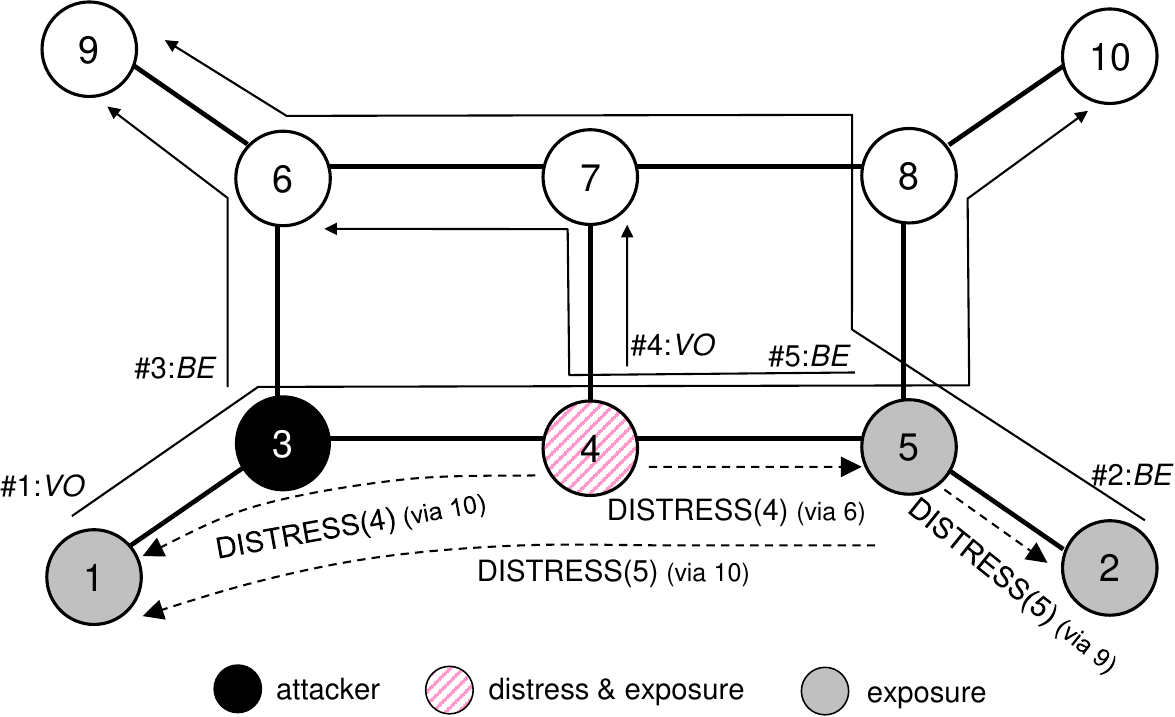}%
\caption{Illustration of the DISTRESS mechanism; e2e-flow \#$i$ originates from source node $i$ and has indicated intrinsic AC; node 4 is in distress due to a TRA$^+$ at node 3.}
\label{fig:distress}
\end{figure}
Fig.~\ref{fig:distress} provides an illustration. Here, $R_{F}^* =  \{(1,3), (1,4), (1,5), (1,6), (1,7),  \allowbreak (1,8), (2,4), (2,5),(2,6), (2,7), (2,8),(3,6), \allowbreak (5,4),(5,7)\}$. Following a TRA at node 3 in stage $k-1$, node 4 sends DISTRESS(4) to $s_{\#1} = 1$ via $d_{\#1} = 10$ and to $s_{\#5} = 5$ via $d_{\#1} = 6$; subsequently, node 5 sends DISTRESS(5) to $s_{\#2} = 2$ via $d_{\#2} = 9$ and to $s_{\#1} = 1$ via $d_{\#1} = 10$; the latter flag is ignored by already in-exposure node 1 (to minimize the communication overhead, it could have been suppressed at node 10). At the end of the present stage, $\Delta^*(\{3\}) = \{1,2,4,5\}$.   

We remark that the DISTRESS mechanism is lightweight in terms of the required synchronization and communication overhead (roughly $\mathcal{O}(|R|)$ per node in the worst case). Exposure signaling is triggered asynchronously by nodes in distress based on local QoS perception. The cause of the distress, either a TRA or temporarily harsh traffic conditions (e.g., transmission impairments, frequent collisions, or buffer overflow), does not influence a node's behavior. Such distinction is a troublesome aspect of many known misbehavior mitigation schemes \cite{Konorski2014}, \cite{konorski2019}, because responding to distress caused by exogenous factors may result in punishment of honest nodes. In our solution, exposure signaling can only encourage a node to select $honest$, therefore does not affect the behavior of an already honest node unintentionally causing distress. By processing a received DISTRESS flag as described above, such a node simply acknowledges an objectively existing threat of imminent service suspension and  no  punishment  occurs.  Importantly, exposure signaling is costless, hence performed without incentive calculation, while fake signaling despite satisfactory QoS perception is not beneficial.

The above considerations also indirectly imply that the effectiveness of the DISTRESS mechanism is unaffected by the network traffic volume, in particular background traffic competing with the e2e-flows: under light traffic conditions no node finds itself in distress and the mechanism is not triggered; otherwise DISTRESS signaling simply encourages honest behavior.

\subsection{``Good'' Multistage Strategies}
\label{sec:multistage-strategy}
A multistage strategy prescribes a node which of the two actions ($TRA$ or $honest$) to select in each stage, based on the current history of play.
To specify a desirable course of play under DISTRESS we use two auxiliary definitions. The first describes flows whose source nodes are not forward-reliant on \emph{in-distress} nodes and so are expected to survive a service suspension. The second describes a node that cannot benefit from selecting a different action, given the other nodes' actions.
\begin{definition}
An e2e-flow is called \emph{survivable} if its source node is not in exposure. Let $F^*(A)$ be the set of such flows in the presence of the attacker set $A \subseteq N$.
\label{auxil3}
\end{definition}
\begin{definition}
A node is called \emph{(weakly) best-reply} if it cannot unilaterally improve its nodal cost (\ref{eq:cost-redefined}) by selecting a different action. Let $\Gamma_{cost'}(A)$ be the set of such nodes in the presence of the attacker set $A \subseteq N$. 
\label{auxil4}
\end{definition}
Thus, we have $F^*(A) = \{ (r,ac) \in F | s_r \notin \Delta^*(A) \}$ and $\Gamma_{cost'}(A) = \{ i \in N | cost'_i(A) \le cost'_i(A^{[i]}) \}$, where $A^{[i]} = A \setminus \{i\}$  if $i \in A $ and $A^{[i]} = A \cup \{i\}$ if $i \notin A$. Note that if $\Gamma_{cost'}(A) = N$ then $A$ constitutes a (weak) Nash equilibrium (NE) \cite{Straffin1993} of the one-shot TRA game (\ref{gamemod}).

We now formulate the following postulates:
\begin{itemize}
    \item \textit{Opt-out} -- while ill-behaved nodes may occasionally select $TRA$, well-behaved nodes select $honest$ at all times; that is, being non-selfish, refuse to play the game. Thus $A(k) \subseteq I$ for all $k$.
    
    \item \textit{Termination} -- after $k_0$ stages the game terminates and no node thereafter changes its action. That is, $A(k) = A(\infty)$ for all $k \ge k_0$, where $k_0$ is finite, known in advance, and preferably small. 

    \item \textit{Rationality} -- ill-behaved nodes tend to select (weakly) best-reply actions; ideally, $I \subseteq \Gamma_{cost'}(A(\infty))$,
    i.e., $A(\infty)$ is a weak NE of the one-shot game restricted to the ill-behaved nodes. A suitable quantitative measure is the fraction of ill-behaved nodes that are best-reply, i.e., $|\Gamma_{cost'}(A(\infty))|/|I|$.

    \item \textit{Efficiency} -- ill-behaved nodes eventually cause one another no distress; ideally, $\Delta(A(\infty)) \cap I = \varnothing$. A suitable quantitative measure is the fraction of ill-behaved nodes that are not in distress, i.e., $|I \setminus \Delta(A(\infty))|/|I|$.

    \item \textit{Defensibility} -- well-behaved nodes are eventually defended against distress caused by ill-behaved nodes' TRAs; ideally, $\Delta(A(\infty)) \subseteq I$. A suitable quantitative measure is the fraction of well-behaved nodes that are not in distress, i.e., $|(N \setminus I) \setminus \Delta(A(\infty))|/|N \setminus I|$.

    \item \textit{Survivability} -- eventually, few e2e-flows rely upon forwarding by nodes in exposure. A suitable quantitative measure of survivable network throughput is the fraction of survivable flows, i.e., $|F^*(A(\infty))|/|F|$.

\end{itemize}

To satisfy these postulates, a multistage strategy has to employ well-designed \textit{action selection} and \textit{participation} rules. The DISTRESS mechanism enables more informed rules by enriching the history of the play: apart from recent actions, a node may recall in-distress conditions and received DISTRESS flags (i.e., in-exposure conditions) in recent stages. We confine a node's memory to the last two stages; the simulations in Section~\ref{sec:simulations} indicate that it can produce satisfactory quantitative measures related to the above postulates. 

We allow node $i$ selecting an action for stage $k + 1$ to recall its membership in the sets of current attackers, current in-distress nodes, and recent in-exposure nodes, i.e., $A(k)$, $\Delta(A(k))$, and $\Delta^*(A(k - 1))$, without the knowledge of the entire sets that the DISTRESS mechanism does not guarantee. Subject to this restriction, a wide class of feasible action selection rules can be expressed as follows:
\begin{equation}
A(k + 1) = \sigma_{A(k),\Delta(A(k)),\Delta^*(A(k - 1))}, 
\label{eq:action-selection}
\end{equation}
where
$$
\sigma_{X,Y,Z} = \bigcup_{(x,y,z) \in \Phi \subseteq \{-1,1\}^3} X^x \cap Y^y \cap Z^z,
$$
$$
\forall X \subseteq N: X^1 = X, X^{-1} = N \setminus X.
$$
Determined by the index set $\Phi$, there are $2^8 = 256$ distinct action selection rules. E.g., if $\Phi = \{(-1,1,-1)\}$ then $\sigma_{X,Y,Z} = X^{-1} \cap Y \cap Z^{-1}$, hence $A(k + 1) = \Delta(A(k)) \setminus (A(k) \cup \Delta^*(A(k - 1)))$, and if $\Phi = \{(-1,1,-1), (1,1,-1)\}$ then $\sigma_{X,Y,Z} = (X^{-1} \cap Y \cap Z^{-1}) \cup (X \cap Y \cap Z^{-1}) = Y \setminus Z$, hence $A(k + 1) = \Delta(A(k)) \setminus \Delta^*(A(k - 1))$. Note that $\Phi = \varnothing$ and $\Phi = \{-1,1\}^3$ correspond to ``persistent honest'' and ``persistent TRA'' strategies, respectively, and that (\ref{eq:action-selection}) subsumes action selection rules with a reduced set of arguments, e.g., $\Phi = \{(-1,y,z),(1,y,z) | (y,z) \in \Phi'\}$, where $\Phi' \subseteq \{-1,1\}^2$, corresponds to an action selection rule that does not explicitly condition $A(k + 1)$ on $A(k)$, i.e., of the form $A(k + 1) = \sigma_{\Delta(A(k)),\Delta^*(A(k - 1))}$.

Participation in the game can change stage by stage as governed by some in-game condition; let $G(k)$ be the set of in-game nodes in stage $k$ that select action according to (\ref{eq:action-selection}), whereas the rest, i.e., $G^{-1}(k)$, retain the previous-stage action. In line with the opt-out postulate, action selection rules and in-game conditions only apply to ill-behaved nodes, i.e., $G(k) \subseteq I$, which we do not reflect in the ensuing formulae to keep them simple.
The dynamics (\ref{eq:action-selection}) become:
\begin{equation}
A(k + 1) = I \cap (G^{-1}(k) \cap A(k)) \cup (G(k) \cap \Sigma),
\label{eq:dynamics}
\end{equation}
where $\Sigma=\sigma_{A(k),\Delta(A(k)),\Delta^*(A(k-1))}$. 
If (\ref{eq:dynamics}) is a deterministic finite-order recurrence then the sequence $(A(k), k = 0,1,2,\ldots)$ eventually becomes periodic and detection of this may terminate the game. It is enough to formulate the in-game condition for a given node that only depends on its recent membership in $A$, $\Delta(A)$, and $\Delta^*(A)$.
Specifically, let $\mathbf{h}_i(k) = (\mathbbm{1}_{i \in A(k-1)},\mathbbm{1}_{i \in A(k)})$ be node $i$ \textit{membership history} with respect to $A(k-1)$ and $A(k)$.
We formulate the following in-game condition for node $i$ in stage $k + 1$:
\begin{equation}
\forall 1 \le c \le \min{\{c_{max}, k - 1\}}: \mathbf{h}_i(k) \neq \mathbf{h}_i(k - c).
\label{eq:history}
\end{equation}
That is, a node is out-of-game if its membership history has repeated itself within recent $c_{max}$ stages (since well-behaved nodes stay honest at all times, they formally become out-of-game as of stage $k = 1$).

We now show that under certain conditions the game is guaranteed to terminate.
\begin{proposition}
\label{prop:1}
If $c_{max} \ge 4$ then termination is guaranteed with $A(k) = A(\infty)$ and $G(k) = \varnothing$ for all $k \ge 8$.
\end{proposition}

\begin{proof}
Observe that out-of-game nodes repeat previous-stage actions, thus if $i \notin G(k)$ and $i \notin G(k + 1)$ then $i \notin G(k')$ for all $k' \ge k$. This is due to $\mathbf{h}_i(k) = \mathbf{h}_i(k + 1)$ being true for all $k' \ge k$, in violation of (\ref{eq:history}) with $c = 1$.

For any node $i \in I$ consider an infinite sequence $\mathbf{a} = (a_0,a_1,a_2,\ldots)$, where $a_0 = 1$ and $a_k = \mathbbm{1}_{i \in A(k)}$ for $k>0$, inducing a sequence of node membership histories $(a_0,a_1), (a_1,a_2), \ldots$ Call $\mathbf{a}$ \textit{proper} if it is compatible with (\ref{eq:history}), i.e., if $\exists 1 \le c \le \min{\{c_{max}, k - 1\}}: (a_k = a_{k-c}) \land (a_{k-1} = a_{k-1-c})$ then $a_{k+1} = a_k$ (hence, $G(k + 1) = \varnothing$). Let $\Pi$ be the set of all proper sequences. Denote $k_0(\mathbf{a}) = \max{\{k | a_k \neq a_k-1\}}$ and, with a little abuse of notation, $k_0(S) = \max{\{k_0(\mathbf{a}) | \mathbf{a} \in S\}}$ for $S \subseteq \Pi$. We need to prove that $k_0(\Pi) = 8$.

Let $\Pi_n$ be the set of all $n$-symbol prefixes of the form $(a_0,\ldots,a_{n-1})$ of sequences in $\Pi$. Based on (\ref{eq:history}) one verifies that if $k_0(\Pi_n) = k_0(\Pi_{n+3})$ then $k_0(\Pi_n) = k_0(\Pi)$.
Inspection shows that among the infinity of sequences $\mathbf{a}$, only 18 are in $\Pi$, and the smallest $n$ satisfying $k_0(\Pi_n) = k_0(\Pi_{n+3})$ is 8, with $k_0(\Pi_8) = k_0(\Pi_{11}) = 8$.
Maximum $k_0(\mathbf{a})$ is achieved at $\mathbf{a^*} = (1,1,0,1,0,0,1,1,0,0,0,0,\ldots)$, inducing a sequence of node membership histories (1,1), (1,0), (0,1), (1,0), \underline{(0,0)}, (0,1), \underline{(1,1)}, (1,0), \underline{(0,0)}, \underline{(0,0)}, \ldots , where the underlined histories correspond to out-of-game stages.
\end{proof}

It turns out that under certain conditions the above specification includes multistage strategies exhibiting ideal rationality and defensibility.
We will need two more definitions. First, we define a nodal cost function under which the benefit of an attacker always causes distress in some other node. Second, we define a flow set such that a service suspension at any node threatens every flow's survival.
\begin{definition}
  The cost function $cost_i(\cdot)$ satisfies \emph{all-honest dominance} if $\varnothing$ is Pareto effective (not Pareto dominated by any $A \subseteq N$), i.e., for all $i \in N$ and $A \subseteq N$, $cost_i(A) < cost_i(\varnothing)$ implies $cost_j(A) > cost_j(\varnothing)$ for some $j \neq i$.
\label{def:domin}
\end{definition}
\begin{definition}
  The flow set $F$ satisfies \emph{full forward-reliance} if $R_{F}^* = N \times N$,
  i.e., each source node is forward-reliant on every other node (cf. Definition~\ref{def:reach}).
\label{def:fullreach}
\end{definition}
In general, all-honest dominance and full forward-reliance are not guaranteed; obviously, the latter is impossible if $R = \{r|(r,ac) \in F\}$ contains single-hop routes. 
However, if both features are present in a given network, then we can show that there exists at least one action selection rule which is ideally efficient (no ill-behaved nodes cause distress), ideally defensible (all well-behaved nodes are defended against distress), and ideally rational (all ill-behaved nodes select best-reply strategies).
\begin{proposition}
\label{prop:2}
If all-honest dominance and full forward-reliance are satisfied then there exists at least one action selection rule $\sigma_{(\cdot)}$ for which:
\begin{enumerate}[label=(\roman*)]
  \item $\Delta(A(\infty)) = \varnothing$, consequently, $\Delta^*(A(\infty)) = \varnothing$ and $cost_i'(A(\infty)) = cost_i(\varnothing)$, i.e., when the multistage game terminates, no node is in exposure and nodal costs are the same as under all-honest (this implies ideal efficiency and defensibility), and
  \item
  $A(\infty)$ is a weak NE of the one-shot game (\ref{gamemod}) restricted to $I$, i.e., $I \subseteq \Gamma_{cost'}(A(\infty))$ (this implies ideal rationality).
\end{enumerate}
\end{proposition}

\begin{proof}
Clearly, all-honest dominance implies $cost_i(\varnothing) \le cost_i'(A)$ for all $A \subseteq N$ and $i \in N$, i.e., with respect to $cost_i'(\cdot)$, $\varnothing$ weakly Pareto dominates every other action profile, and as such is a weak NE of the one-shot game (\ref{gamemod}).
The same pertains to any $A \subseteq N$ such that $\Delta(A) = \varnothing$ (indeed, for any such $A$, $cost_i(A) = cost_i(\varnothing)$ for all $i \in N$).
Since under full forward-reliance it holds that $\Delta(A) = \varnothing$ iff $\Delta^*(A) = \varnothing$, it is enough to show that $\Delta^*(A(\infty)) = \varnothing$.

Let $k-1 = \min\{k'|G(k')= \varnothing\}$, then from (\ref{eq:history}) it follows that $A(k-1) = A(k) = A(k + 1) = A(\infty)$, where $A(\infty) = \sigma_{A(\infty),\Delta(A(\infty)),\Delta^*(A(\infty))}$. If $A(\infty) = \varnothing$ then the proposition trivially holds true, so assume $A(\infty) \neq \varnothing$. Take $\Phi = \{(-1,-1,-1), (-1,1,-1), (1,-1,-1), (1,1,-1)\}$, i.e.,  $\sigma_{X,Y,Z} = Z^{-1}$ and $A(k + 1) = I \cap (N \setminus \Delta^*(A(k - 1)))$. Then either $\Delta(A(\infty)) = \Delta^*(A(\infty)) = \varnothing$ and the proposition holds true, or $\Delta(A(\infty)) \neq \varnothing$, which implies $\Delta^*(A(\infty)) \neq N$, impossible under full forward-reliance.
\end{proof}
Proposition \ref{prop:2} states conditions of existence of ideally rational, efficient, and defensible action selection rules. Simulations show that these conditions are often satisfied in randomly generated MAWiN instances; otherwise, ``good'' rules (\ref{eq:action-selection}) nevertheless exist that ensure satisfactory characteristics across various MAWiN topologies and flow sets $F$, cf. Section~\ref{sec:simulations}. These rules can be adopted\textit{ \'{a} priori}, relieving nodes from seeking ``good'' rules for a specific topology and flow set, which they are typically unaware of.

Note that $\Delta(A) = \varnothing$ implies that $A$ is a weak NE of (\ref{gamemod}), but the converse is not true; in fact, simulations show that a vast majority of Nash equilibria $A$ feature $\Delta(A) \ne \varnothing$.

\section{Simulations}
\label{sec:simulations}
Simulations involved network topologies with both static and mobile nodes, respectively using the Monte Carlo method based on the network model of Section \ref{sec:model} and the ns-3 simulator implementing a full MAWiN protocol stack. The goal of the Monte Carlo simulations was to assess the rationality, efficiency, defensibility, and survivability measures in realistic settings, when the two assumptions of Proposition~\ref{prop:2} (all-honest dominance and full forward-reliance) were not necessarily satisfied. The ns-3 simulations were carried out to investigate the impact of fast-changing MAWiN topology and flow routes on the effectiveness of the DISTRESS mechanism. 
\subsection{Static Nodes}
For the analysis of static topologies, we implemented the network model of Section~\ref{sec:model} (including network topology, routing, flow configuration, attack behavior, and rank-based estimation of flow cost) as well as the multi-stage TRA game and DISTRESS mechanism of Section~\ref{sec:game} in a Monte Carlo simulator.
Each simulation run consisted of a stage-by-stage play of the TRA game, with nodes deciding their action (\textit{TRA} or \textit{honest}) in each stage.
1000 MAWiN instances were generated with $|N| = |F| = 10$ and e2e-flow routes of uniformly distributed hop lengths with $\lVert r \rVert_{\text{min}}$ = 1 or 2 and $\lVert r \rVert_{\text{max}}$ = 5. Full forward-reliance occurred in 6.5\% and 55.2\% of the MAWiN instances with $\lVert r \rVert_{\text{min}}$ = 1 and $\lVert r \rVert_{\text{min}}$ = 2, respectively. We observed that:
\begin{itemize}
  \item All-honest dominance was never violated (violations are in general possible but unlikely, e.g., when traffic flows do not impact one another's performance). 

\item For $\lVert r \rVert_{\text{min}}$ = 1 and $\lVert r \rVert_{\text{min}}$ = 2, weak Nash equilibria amounted to 40.2\% and 94\% of all action profiles $A \subseteq N$, respectively; however, those with $\Delta(A) = \varnothing$ were a rarity, amounting to 0.1\% and 0.3\%, respectively.

\end{itemize}
For each MAWiN instance, 100 independent multistage game runs were conducted with $I \neq N$ chosen at random subject to $\Delta(I) \neq \varnothing$. 
All 256 feasible action selection rules were tried. It was observed that:
\begin{itemize}  
  \item With $\lVert r \rVert_{\text{min}}$ = 1 and $\lVert r \rVert_{\text{min}}$ = 2, respectively 44 and 54 action selection rules
  (including the trivial
  ``persistent honest'') ensured (\textit{i}) and (\textit{ii}) of Proposition~\ref{prop:2} even in the absence of full forward-reliance.
  
  \item Conditioning $A(k + 1)$ on $A(k)$ besides $\Delta(A(k))$ and $\Delta^*(A(k-1))$ did not extend the interesting range of key after-game (asymptotic) characteristics.
  \item On the other hand, conditioning $A(k + 1)$ on $A(k)$ alone, i.e., disregarding information provided by the DISTRESS mechanism, produced about the worst measures of rationality and defensibility. 
\end{itemize}

\begin{figure}[t]
\centering
\includegraphics[width=\columnwidth]{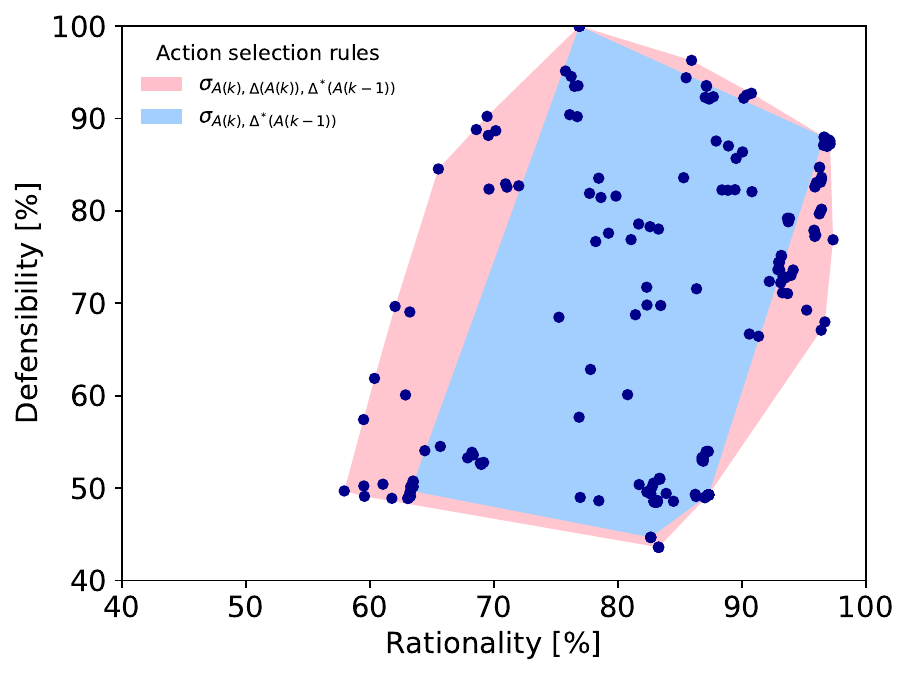}%
\caption{Rationality and defensibility measures for the $2^8$ feasible action selection rules; $|N| = |F| = 10, \lVert r \rVert_{\text{min}}$ = 1.}
\label{fig:d-vs-r}
\end{figure}

To explain the first observation consider, e.g., %
$A(k + 1) = I \cap \Delta(A(k))$. It must be that $A(\infty) = \Delta(A(\infty))$, but $A(\infty) = \Delta(A(\infty)) \neq \varnothing$ is possible only if eventually all the attacker nodes and none of the honest ones are caused distress, a highly improbable situation.
The latter two observations are illustrated in Fig.~\ref{fig:d-vs-r},
where each dot represents a pair of after-game (rationality, defensibility) measures for a given action selection rule, averaged over the generated MAWiN instances and game runs.
The outer contour encompasses all action selection rules (\ref{eq:action-selection}) and the inner one only rules of the form $A(k + 1) = \sigma_{\Delta(A(k)),\Delta^*(A(k - 1))}$. The latter captures most of the Pareto front including the best rationality and defensibility measures. Rules of the form $A(k + 1) = \sigma_{A(k)}$ correspond to the two lower corners of the outer contour, the farthest from the Pareto front.

For more detailed analysis, several representative action selection rules have been chosen. They are listed below in the order of diminishing survivability and tendency to launch TRAs, and improving efficiency and defensibility:
\begin{enumerate}[label=(\alph*)]
\item $A(k + 1) = N$ (``persistent TRA''), 
\item $A(k + 1) = \Delta(A(k)) \cap \Delta^*(A(k - 1))$,
\item $A(k + 1) = \Delta(A(k)) \oplus \Delta^*(A(k - 1))$, where $\oplus$ denotes disjunctive union,
\item $A(k + 1) = (N \setminus \Delta^*(A(k - 1))) \cup \Delta(A(k))$,
\item $A(k + 1) = N \setminus \Delta^*(A(k - 1))$,
\item $A(k + 1) = \Delta(A(k)) \setminus \Delta^*(A(k - 1))$.
\end{enumerate}
These rules were observed to differ visibly in the speed of convergence to the after-game characteristics; example stage-by-stage trajectories for rules (e) and (f) are shown in Fig.~\ref{fig:trajectories}.
Table~\ref{tab:convergence} presents relevant after-game characteristics, averaged as above. One notes in particular that:

\begin{figure}[t]
\centering
\iffalse
\subfloat{\includegraphics[width=0.43\textwidth]{trajectory-a.pdf}%
\label{fig:trajectory-a}}
\subfloat{\includegraphics[width=0.43\textwidth]{trajectory-b.pdf}%
\label{fig:trajectory-b}}\\
\subfloat{\includegraphics[width=0.43\textwidth]{trajectory-c.pdf}%
\label{fig:trajectory-c}}
\subfloat{\includegraphics[width=0.43\textwidth]{trajectory-d.pdf}%
\label{fig:trajectory-d}}\\
\fi
\subfloat{\includegraphics[width=0.43\textwidth]{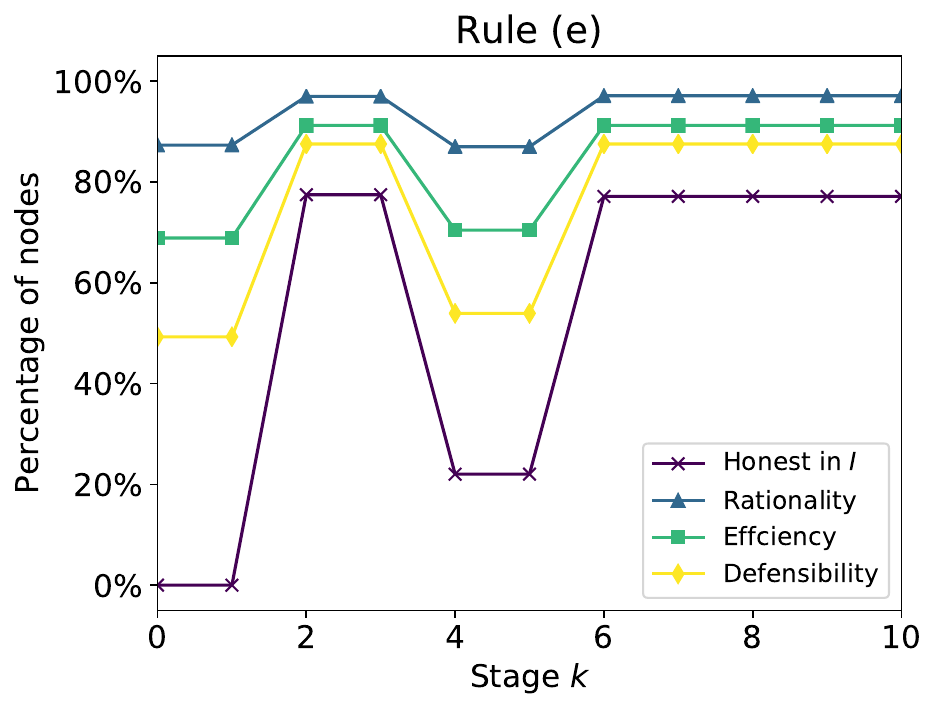}%
\label{fig:trajectory-e}}\\
\subfloat{\includegraphics[width=0.43\textwidth]{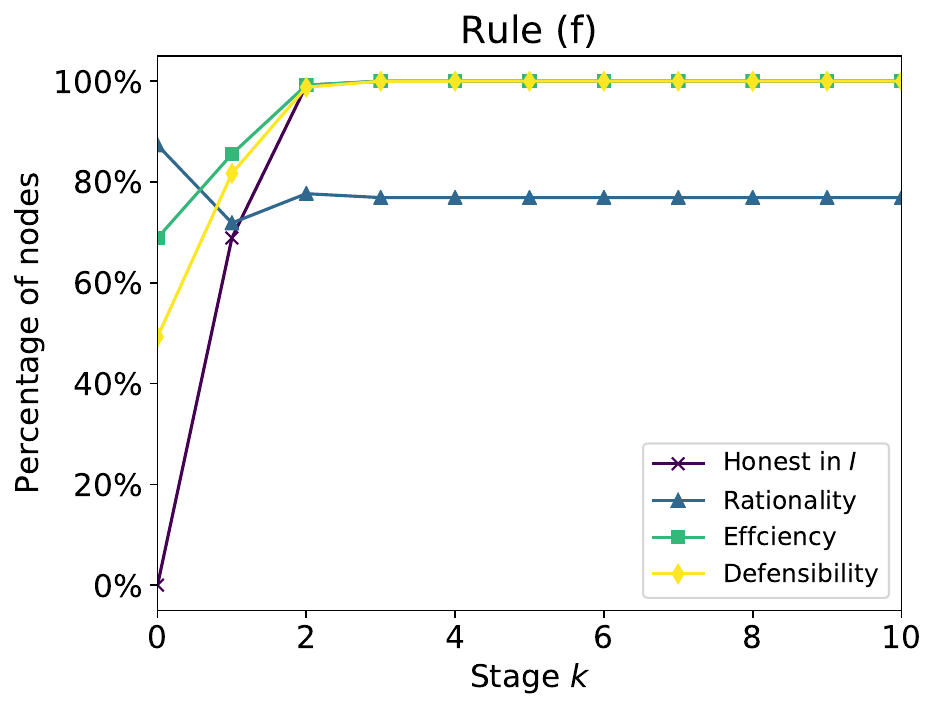}%
\label{fig:trajectory-f}}
\caption{Stage-by-stage trajectories of relevant characteristics for action selection rules (e) and (f); $\lVert r \rVert_{\text{min}}$ = 1.
}
\label{fig:trajectories}
\end{figure}	

\begin{table*}[t]
\begin{threeparttable}
\caption{Simulation results: average after-game characteristics for action selection rules (a) to (f); $\lVert r \rVert_{\text{min}} = 1$.}
\footnotesize
\label{tab:convergence}
\begin{tabularx}{\linewidth}{XXXXXXX}
\toprule
Action selection rule & Honest in $I$ [\%] & Rationality [\%]$^*$
& Efficiency [\%]
& Defensibility [\%]
& Survivability [\%]
& Game duration [stages] \\ \midrule
(a)                   & 0                     & 87.3 (79.4)                                 & 68.9                                   & 49.3                                       & 26.5                                     & 3                      \\
(b)                   & 31.1                    & 82.9 (66.8)                                 & 72.9                                   & 48.6                                       & 29.3                                     & 3.8                    \\
(c)                   & 42.6                    & 64.4 (37.0)                                 & 76.6                                   & 54.0                                         & 36.9                                     & 4.6                    \\
(d)                   & 49.0                      & 93.2 (79.2)                                  & 81.5                                   & 75.2                                       & 50.7                                     & 4.9                    \\
(e)                   & 77.1                    & 97.1 (90.7)                                  & 91.2                                    & 87.4                                       & 73.3                                     & 5.1                    \\
(f)                   & 100                       & 76.9 (48.6)                                 & 100                                      & 100                                          & 100                                        & 3.7                    \\ \bottomrule
\end{tabularx}
\begin{tablenotes}
  \footnotesize
  \item [*] In parentheses are percentages of game runs leading to Nash equilibria within $I$.
\end{tablenotes}
\end{threeparttable}
\end{table*}

\begin{itemize}
    \item Rule (e) is the most likely to be adopted by rational ill-behaved nodes, as it leads to highly efficient Nash equilibria in the vast majority of game runs. It also produces good defensibility and survivability.

    \item On the other hand, rule (f) produces ideal efficiency and survivability, but is far from rational, thus likely to be dismissed by ill-behaved nodes.

    \item Rule (a) (``persistent TRA'') is moderately rational, but, due to the DISTRESS mechanism, very inefficient. This latter fact is fortunate, as rule (a) produces disastrous defensibility and survivability.

    \item Rules adversely affecting defensibility and survivability, such as (a) through (c), are not very rational and so unattractive to ill-behaved nodes. Rule (d) is moderately attractive, but from rationality and efficiency viewpoints is Pareto dominated by rule (e).
\end{itemize}
Hence, under DISTRESS, ill-behaved nodes are likely to follow rule (e), in which in-exposure nodes cannot be attackers.

\subsection{Mobile Nodes}

\begin{table}[t]
\centering
\begin{threeparttable}
\caption{ns-3 simulation parameters}
\footnotesize
\label{tab:ns3params}
\begin{tabular}{@{}ll@{}}
\toprule
Parameter             & Value           \\ \midrule
Total number of nodes & 16              \\
Node placement        & grid, random    \\
Area size             & 20 x 20 m       \\
Grid spacing          & 8 m             \\
Mobility model        & random waypoint \\
Waiting time          & 20 s            \\
Velocity              & 0 -- 1.4 m/s    \\
PHY/MAC               & IEEE 802.11ac   \\
RTS/CTS               & Enabled         \\
Channel width         & 20 MHz          \\
Modulation            & 256-QAM         \\
Coding rate           & 3/4             \\
aCWmin                & 63\tnote{1}     \\
Routing               & OLSR\tnote{2}  \\
Transport protocol    & UDP             \\
Traffic generator     & CBR             \\
Traffic rate          & 2 Mb/s          \\ \bottomrule
\end{tabular}
\begin{tablenotes}
  \footnotesize
  \item [1] To  reduce the effect of hidden nodes.
  \item [2] OLSR's signaling messages were prioritized over VO flows.
\end{tablenotes}
\end{threeparttable}
\end{table}

To study the impact of node mobility, we
implemented the multi-stage play of Section~\ref{sec:multistage-strategy}  under rule (e) in the ns-3 simulator, using the settings of Table~\ref{tab:ns3params}.
Each node was the source of an e2e-flow, half of them of high priority.
We considered two topologies for initial node placement: \emph{grid} -- all nodes arranged on a square grid, and \emph{random} -- all nodes uniformly distributed in the area. 
The adopted high modulation and coding scheme limits the nodes' transmission range to approximately 10~m. For a more realistic IEEE 802.11 range of 100~m, the maximum evaluated velocity would scale to 14~m/s (50~km/h).
In summary, the chosen settings created conditions when TRAs are likely to negatively impact the performance of honest nodes.

With node mobility, $\Delta(A)$ and $\Delta^*(A)$ can change stage by stage although $A$ remains the same, due to changing MAWiN topology and flow routes. Except for quasi-static environments, where typical TRA game duration (on order of a few stages) fits between successive route changes, this may complicate ill-behaved nodes' strategic behavior. For preliminary insight we assumed that they nevertheless stick to rule (e); in particular, they calculate $cost_i(\varnothing)$ only once for distress perception, at game start, and maintain the current action (TRA or honest) indefinitely when the out-of-game condition is satisfied. The key question was whether the DISTRESS mechanism would still be able to restrain TRAs.
As an evaluation metric for our DISTRESS mechanism, we used the percentage of after-game attackers, which reflects how many of the network's ill-behaved nodes remain attackers after the game has terminated. Using this metric, we compared the performance of a MAWiN under the DISTRESS mechanism with a baseline MAWiN, where nodes do not fear service suspension and remain attackers if such behavior improves the QoS metrics  (\ref{cost})  of their source flows.

\begin{figure*}[t]
\centering
\subfloat[]{\includegraphics[width=0.5\textwidth]{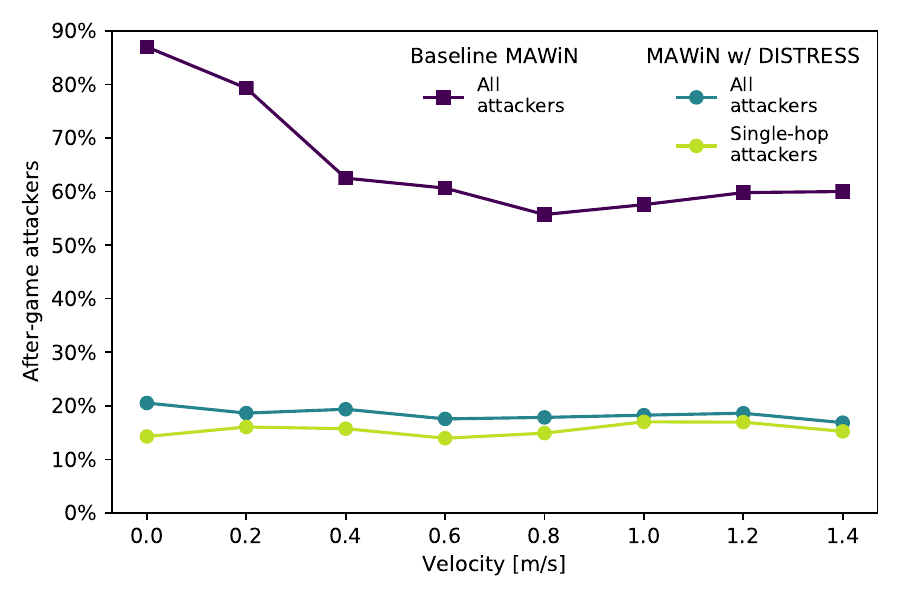}%
\label{fig:grid-topology}}
\subfloat[]{\includegraphics[width=0.5\textwidth]{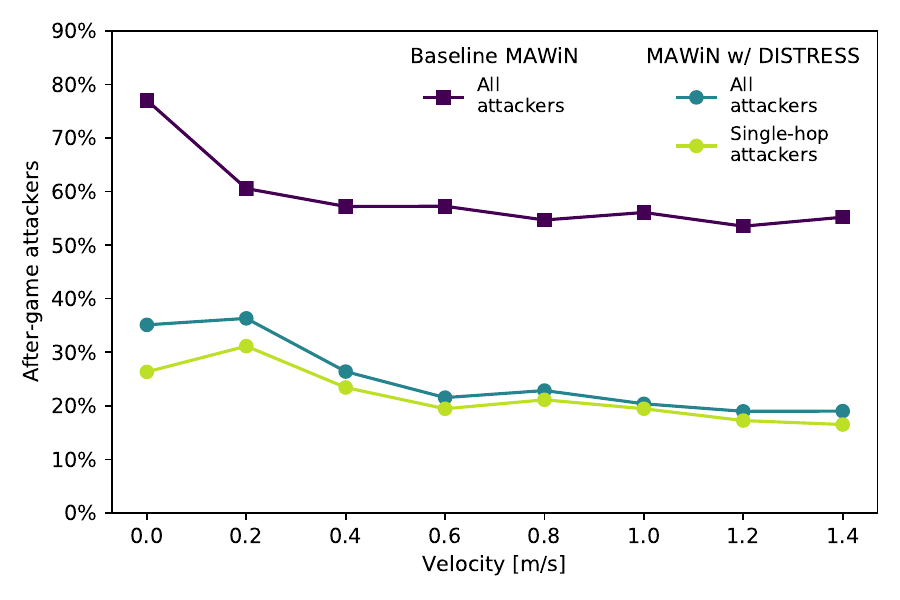}%
\label{fig:random-topology}}
\caption{
Percentage of after-game attackers in $N$ for varying node velocities in the case of grid (a) and random (b) initial topologies.
}
\label{fig:mobility}
\end{figure*}	

In Fig.~\ref{fig:mobility} each point is an average of 100 independent simulation runs with half the nodes being ill-behaved ($|I| = |N|/2$). Simulations confirmed that even for the maximum node velocity the adopted settings created a quasi-static environment -- route changes during the TRA game were occasional or none and the play closely resembled that of Section~\ref{sec:multistage-strategy}. Hence, irrespective of node velocity and initial placement, after-game attackers were indeed distinctly fewer than for the baseline MAWiN with the DISTRESS mechanism disengaged; they were mostly limited to ill-behaved sources of single-hop flows, which, having no forwarding services to rely upon, remained unaffected by the DISTRESS mechanism.

\begin{figure}[t]
\centering
\includegraphics[width=\columnwidth]{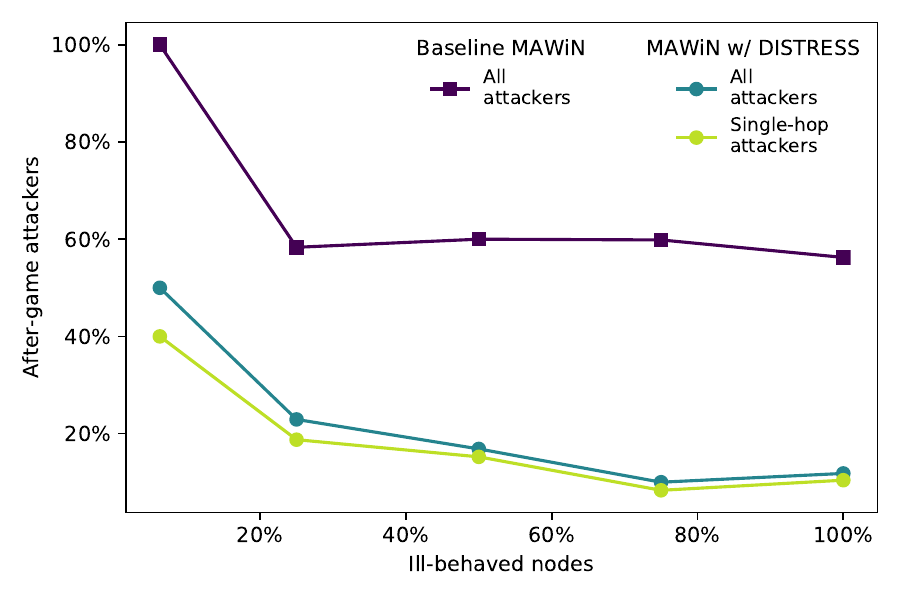}%
\caption{
Percentage of after-game attackers in $N$ for varying percentage of ill-behaved nodes in the grid topology with high mobility.
}
\label{fig:attackers-grid}
\end{figure}	

\begin{figure}[t]
\centering
\includegraphics[width=0.9\columnwidth]{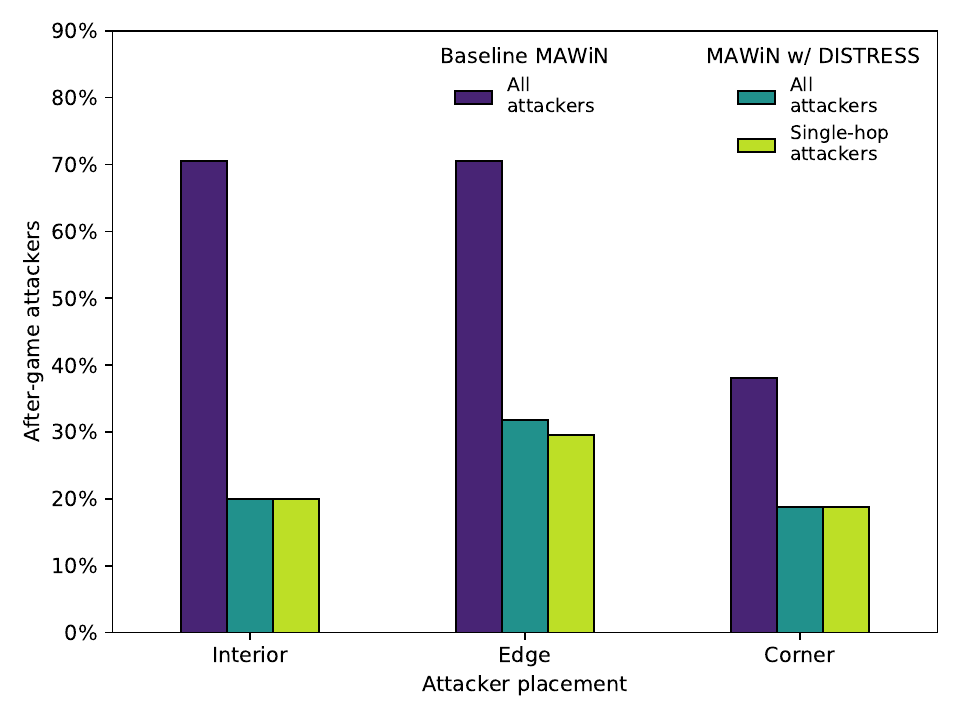}%
\caption{
Percentage of after-game attackers depending on initial ill-behaved node placement in the grid topology with high mobility.
}
\label{fig:attacker-placement}
\end{figure}	

To evaluate our proposed mechanism against a varying attack intensity, we note that Definitions \ref{def:intro-tra}-\ref{def:plausible} do not leave room for any gradation of TRA intensity: a node either behaves honestly or executes a plausible opportunistic TRA. Therefore, overall attack intensity in the network is sufficiently reflected by the percentage of ill-behaved nodes in $N$. Fig.~\ref{fig:attackers-grid} presents the respective simulation results for a high-mobility (1.4 m/s) scenario of the grid topology. DISTRESS is uniformly able to incentivize honest behavior in all (or almost all) of those ill-behaved nodes which are the sources of multi-hop flows. As before, only the sources of single-hop flows remain attackers.

Finally, even though a node's placement in the network topology does influence the potential benefits of becoming an attacker (as is visible in Fig.~\ref{fig:topology}), one suspects that the effectiveness of the DISTRESS mechanism does not depend on the placement of attackers. This is because DISTRESS signaling is network-wide and affects the cost metric (\ref{gamemod}) of each node regardless of its location. Fig.~\ref{fig:attacker-placement} shows that indeed the initial placement (interior, edge, or corner) of ill-behaved nodes in a grid-topology MAWiN does not impact the percentage of after-game attackers.

\section{Conclusion}
\label{sec:conclusions}
A traffic remapping attack (TRA) is hard to defend against in MAWiNs due to their multi-hop topology, node autonomy, and complex interplay of factors affecting end-to-end performance.
We have proposed a systematic game-theoretic approach to TRA mitigation.
The adopted model of a MAWiN under plausible opportunistic TRAs allows to define a noncooperative multistage TRA game arising among selfish nodes, in which the payoff function is provided by a novel network-oriented end-to-end QoS metric. We have augmented this function to reflect the threat of forwarding service suspension due to ongoing TRAs, as disseminated by a robust and low-cost distributed signaling mechanism called DISTRESS.

Our work distinguishes itself by proposing the first distributed self-enforcing mitigation approach for TRAs in MAWiNs. Existing solutions are not directly comparable: they counteract other types of attacks, rely on attack detection, require centralized control, or have been designed for single-hop networks and cannot cope with the multi-hop nature of TRAs.
However, an advantage of the proposed framework is that by analyzing all feasible action selection rules (\ref{eq:action-selection}), it enables an exhaustive search of a wide class of selfish nodes' multistage strategies. Therefore, optimum rules can be found according to the selected criteria, so that  comparisons with particular solutions, existing or to be found in the future, are less relevant.

Our framework also enables a precise statement of postulates regarding a desirable game outcome: opt-out (well-behaved nodes need not play), termination (finite game duration is guaranteed), rationality (ill-behaved nodes select best-reply behavior), defensibility and efficiency (well- and ill-behaved nodes receive satisfactory QoS), and survivability (little traffic is threatened by forwarding service suspension). We have argued that ill-behaved nodes are likely to use a strategy which, under certain assumptions regarding MAWiN topology and traffic flows, guarantees that these postulates are satisfied. The game outcome remains desirable even for a broader class of static-topology MAWiNs, as demonstrated by Monte Carlo simulations; time-true simulations using ns-3 extend this conclusion to networks with mobile nodes.

Although it has been analyzed assuming fixed e2e-flow routes, the DISTRESS mechanism can work with alternate routing as well. Provided that a node is aware of all currently available routes for the originated e2e-flow, it only needs to mark itself as in-exposure upon reception of a DISTRESS flag from at least one node on each of these routes. Under dynamic routing, the forward-reliance relationship may be time-varying; a DISTRESS flag received from a node on a given route can then be interpreted by a source node as a noncommittal signal to propagate DISTRESS flags further, depending on the projected route stability.

Finally, mechanisms providing QoS security, similar to DISTRESS, might produce viable game-theoretic defense against QoS abuse in other distributed settings offering QoS differentiation; this is left for future research.

\section*{Acknowledgments}

The work of Jerzy Konorski is funded by the National Science Center, Poland, under Grant UMO-2016/21/B/ST6/03146. 
The work of Szymon Szott is supported by the Polish Ministry of Science and Higher Education with the subvention funds of the Faculty of Computer Science, Electronics and Telecommunications of AGH University.
This research was supported in part by PLGrid Infrastructure.

\bibliographystyle{IEEEtran}
\bibliography{IEEEabrv,tra-game}

\end{document}